\documentclass[12pt, fleqn]{article}
\usepackage[cp1251]{inputenc}
\usepackage{latexsym,amsfonts,amssymb}
\usepackage{graphicx}
\usepackage{amsbsy}
\usepackage{amsmath}
\usepackage{epsf}
\usepackage{cite}

\newcommand{\be} {\begin{equation}}
\newcommand{\ee} {\end{equation}}
\newcommand{\ba}{\begin{array}{l}}
\newcommand{\ea}{\end{array}}

\newtheorem{theo}{Theorem}
\newtheorem{remark}{Remark}
\newtheorem{consequence}{Consequence}
\newcommand{\bteo}{\begin{theo}}  
\newcommand{\et}{\end{theo}}

\newcommand{\bn}{\begin{note}}
\newcommand{\en}{\end{note}}

\begin{document}

\begin{center}
 {\Large \bf Lie symmetries  \\
 of nonlinear  parabolic-elliptic   systems \\ and their application to a tumour growth model}
 \medskip

 {\bf Roman Cherniha,$^{a}$\footnote{\small  Corresponding author. E-mail: r.m.cherniha@gmail.com}}
  {\bf Vasyl' Davydovych$^a$\footnote{\small  e-mail: davydovych@imath.kiev.ua}} and
 {\bf John R. King$^b$\footnote{\small e-mail: John.King@nottingham.ac.uk}}
 \\
{\it
$^a$~Institute of Mathematics,  National Academy
of Sciences  of Ukraine,\\
 3, Tereshchenkivs'ka Street, Kyiv 01601, Ukraine\\
  $^b$~\it School of Mathematical Sciences,
University of Nottingham, \\
University Park, Nottingham, NG7 2RD, UK}
 \end{center}
\textbf{Abstract.} A generalisation of the  Lie symmetry method is
applied to classify a coupled system of reaction-diffusion equations
wherein the nonlinearities involve arbitrary functions in the limit
case in which one equation of the pair is quasi-steady but the other
not. A complete Lie symmetry classification, including  a number of
the cases characterised being   unlikely   to be identified purely
by intuition, is obtained. Notably, in addition to the symmetry
analysis of the PDEs themselves, the approach is extended to allow
the derivation  of exact solutions to specific
 moving-boundary problems motivated by biological applications (tumour growth). Graphical
representations of the solutions are provided and biological
interpretation addressed briefly.
The results   are generalised on
multi-dimensional case  under assumption of radially
  symmetrical shape of the tumour.

 \section{\bf Introduction}

It is  well-known that  nonlinear reaction-diffusion systems  (i.e.
systems of second-order parabolic  PDEs with reaction terms) are
used to describe various processes in physics, chemistry, biology,
ecology, etc (see, e.g.,  books
 \cite{ames, mur2, mur2003, okubo}). Indeed, there are many special cases of such
  systems describing real world  processes, among   the most famous being  Lotka--Volterra type
  systems \cite{lotka,volterra} and systems used for
 modelling   the chemical basis of morphogenesis \cite{turing} (see, e.g., \cite{mur2} for
 details in modern terminology).

 The most common are two-component reaction-diffusion systems
of the form
  \begin{equation}\label{1-1}
 \begin{array} {l}
U_t = (D_1(U)U_{x})_{x}+F(U,V),\\
V_t = (D_2(V)V_{x})_{x}+G(U,V),
 \end{array}
  \end{equation}
where $U(t,x)$ and $V(t,x)$ are unknown functions,  the nonnegative
smooth functions $D_1(U)$ and $D_2(V)$ are coefficients of
diffusivity (conductivity), and $F$ and $G$ are arbitrary smooth
functions describing interaction between components $U(t,x)$ and
$V(t,x)$.  Hereafter the $t$ and $x$ subscripts denote
differentiation with respect to these variables.

It should be noted that the search of Lie symmetries of  RD systems
was initiated many years ago and probably    paper
\cite{zulehner-ames} was the first in this direction. A complete Lie
symmetry
 classification was done much later because this problem is
  much more complicated in comparison with the case of a single reaction-diffusion equation.
  In particular, it was necessary to treat separately the cases of constant
   diffusivities (i.e. semilinear equations) and nonconstant (quasilinear) ones.
All possible  Lie symmetries of  (\ref{1-1}) with constant
diffusivities
 were  completely described in
\cite{ch-king,ch-king1,ch-king2}.  In the case of  non-constant
diffusivities,
 it has been done in  \cite{ibrag-94, ch-king4} (in \cite{ch-king06},
  the result was extended to the multi-component RD systems).

It should be stressed that there are models used to describe real
world  processes that are based on singular limits of a RD system
(\ref{1-1}), namely with $D_2(V)=0 $ or with the $V_t$ term absent.
The system (\ref{1-1}) with $D_2(V)=0 $ has been used to describe
interactions involving species (cells) $V$, which are not able to
diffuse in space (typical examples include some models for
plankton). Some particular results about Lie symmetries of such
systems have been recently  obtained in \cite{tor-tra-15}, however
the problem of  a complete Lie symmetry classification is still not
solved.

Here, however,  we examine the class of systems (\ref{1-1}) with the
$V_t$ term negligible so that $V$ is governed by a quasi-steady
equation, i.e.
 \begin{equation}\label{1-2}
 \begin{array} {l}
U_t = (D_1(U)U_{x})_{x}+F(U,V),\\
0 = (D_2(V)V_{x})_{x}+G(U,V).
 \end{array}
  \end{equation}
Obviously that this system is reduced to the form
 \begin{equation}\label{1-3}
 \begin{array} {l}
U_t = (D_1(U)U_{x})_{x}+F^*(U,V^*),\\
0 = V^*_{xx}+G^*(U,V^*)
 \end{array}
  \end{equation}
  by the Kirchhoff substitution for the component $V \to V^*$. Thus, the class of parabolic-elliptic  systems (\ref{1-3}) is the main object of this study.
The main motivation for this follows from paper \cite{by-ki-2003},
in which a model for
  tumour growth was derived. Under some biologically motivated
  assumptions
   (see p.570 in \cite{by-ki-2003} for details)   the model was simplified
    to a boundary-value problem with governing equations of the form
    (\ref{1-3}). System  (\ref{1-2}) is of course  potentially of
    relevance to any two-component reaction-diffusion model in
    which one species diffuses much faster than the other; it is
    also of interest purely from the symmetry point of view, since
    its symmetries differ essentially from those of (\ref{1-1}).

The paper is organized as follows. Section 2 is devoted to search
for form-preserving transformations for the class of systems
(\ref{1-3}) in order to establish possible relations between RD
systems that  admit equivalent  Lie symmetry algebras. Section 3 is
devoted to the identification of  all possible Lie symmetries that
any system of the form (\ref{1-3})  can admit. The main results of
this section are presented in the form of Tables~\ref{tab-1}
and~\ref{tab-2}. In Sections 4 and 5, we apply the  Lie symmetries
derived for reduction of a nonlinear  boundary value problem (BVP)
with a free boundary modeling tumour growth in order to construct
its exact solution. Moreover, possible  biological interpretations
of the results derived is discussed.  Finally, we briefly discuss
the result obtained and present some conclusions in the last
section.

\section{Form-preserving transformations for the class of systems (\ref{1-3}) }

First of all we simplify notations in  systems (\ref{1-3}) in a natural way:
 \begin{equation}\label{1a}
 \begin{array} {l}
 U_t = (D(U)U_{x})_{x}+F(U,V),\\
0 = V_{xx}+G(U,V).
 \end{array}\end{equation}

The class of systems (\ref{1a}) contains three arbitrary functions
and the Lie symmetry of its different representatives depends
essentially
 on the form the triplet $(D,F,G)$. Thus, the problem of a
complete description of all possible Lie symmetries (the so called
group classification problem) arises. In order to solve this, we can
apply the  Lie--Ovsiannikov  approach (the name of  Ovsiannikov
arises because he  published a remarkable paper in this direction,
\cite{ovs-1959}) of the Lie symmetry classification, which  is based
on the classical Lie scheme and a set of equivalence transformations
of  the  differential equation in question.
However, it is well-known that  this approach leads to very long list of equations with non-trivial Lie
 symmetry provided the given equation (system) contains several arbitrary functions
  (system (\ref{1a}) involves three such).

During the last two decades  new approaches  for solving  group classification problems
 were developed,  which  are  important for obtaining the  so called canonical
  list of inequivalent equations admitting non-trivial Lie symmetry algebras and allow the solution of  this  problem
 in a more efficient way than a  formal application of the Lie--Ovsiannikov approach.
Here we use the algorithm based on so called  form-preserving
transformations \cite{kingston-91,kingston-98},
which were used initially for finding locally-equivalent PDEs,
especially those nonlinear  PDEs that are linearizable by a point
transformation.
 Interestingly   such transformations were implicitly used much earlier in
 \cite{niederer-78} in order to find  all possible heat equations with nonlinear sources
  that admit the  Lie symmetry either of the linear heat equation or the Burgers
  equation. In \cite{winternitz1992}, these transformations are called
   `admissible transformations' and they were used to classify the Lie symmetries
 of a class of variable  coefficient  Korteweg-de  Vries  equations.

 It was noted later (probably, paper \cite{ch-se-98} was the first in this direction)
  that the form-preserving transformations (other terminology is `additional equivalence
   transformations' or `admissible transformations')
  allow an essential  reduction of  the number of cases obtained via the classical Lie--Ovsiannikov algorithm
   (see   extensive discussions  on this matter  in  \cite{ch-king2,ch-king4,ch-se-ra-08}).
For example, it was  proved in \cite{ch-se-ra-08} using a set of
form-preserving transformations that the canonical list of
inequivalent reaction-diffusion-convection equations consists of 15
equations only (not the 30 derived by the Lie--Ovsiannikov
algorithm).

Now we present the main result of this section, namely
 the theorem describing a general form of form-preserving transformations for the  class of systems (\ref{1-3}).

\begin{theo}\label{th-1} An arbitrary   system of the form (\ref{1a})
be reduced to another   system of the same form
   \begin{equation}\label{2-29}\begin{array}{l}
W_{\tau}=\left(\lambda(W)W_y\right)_y+\tilde{F}(W,Z),
\\ 0=Z_{yy}+\tilde{G}(W,Z)\end{array}\end{equation}
by the  local non-degenerate
transformation
\begin{eqnarray}
& &\tau=a(t,x,U,V), \ y=b(t,x,U,V),  \nonumber\\
& & W=\varphi(t,x,U,V), \ Z=\psi(t,x,U,V),\nonumber
\end{eqnarray}
 if and only if the smooth functions $a, \ b, \ \varphi$ and  $\psi$ have the form
\begin{equation}\label{2-30}\begin{array}{l} a=\alpha(t),  \ b=\beta(t,x), \qquad \quad \dot{\alpha}\beta_x\neq0,\\
\varphi=K(t,x)U+P(t,x), \quad K\neq0,\\
\psi=L(t,x)V+Q(t,x), \quad L\neq0, \end{array}\end{equation} where
the functions  $\alpha(t), \ \beta(t,x), \ K(t,x), \ P(t,x), \
L(t,x)$ and  $Q(t,x)$ are such that  the following equalities:
\begin{eqnarray} &&
\dot{\alpha}\lambda=\beta^2_xD, \nonumber \\
&& \dot{\alpha}\tilde{F}=KF+\frac{(K_xU+P_x)^2}{K}\frac{\partial
D}{\partial
U}+K_tU+P_t-D\left(K_{xx}U+P_{xx}-2\frac{K_xU+P_x}{K}K_x\right), \nonumber \\
&& \beta^2_x\tilde{G}=LG+2\frac{L_xV+Q_x}{L}L_x-L_{xx}V-Q_{xx},
\label{2-33} \\
&& 2\beta_x(K_xU+P_x)\frac{\partial D}{\partial
U}+\left(2\beta_xK_x-\beta_{xx}K\right)D+\beta_tK=0, \nonumber \\
&& 2\beta_xL_x-\beta_{xx}L=0,\nonumber
\end{eqnarray}
hold. \end{theo}

\textbf{The proof of Theorem~\ref{th-1}}  is quite similar to that
for the class reaction-diffusion systems  (\ref{1-1}) with
$D_k=constant \ (k=1,2)$ presented in \cite{che-dav2014}.


The group of equivalence transformations can be easily extracted
from Theorem~\ref{th-2} by assuming that  $D,F$ and $G$ are
arbitrary smooth functions. Solving the system  (\ref{2-33}) under
such condition one arrive at the following statement.

\begin{consequence} The group of equivalence transformations for
(\ref{1a}) has the form
\begin{equation}\label{2-36}\begin{array}{l} \tau= \alpha_1t+\alpha_2,\quad y= \alpha_3x+\alpha_4, \\
w= \alpha_5u+\alpha_6,\quad z= \alpha_7v+\alpha_8,\\
\lambda=\frac{\alpha^2_3}{\alpha_1}D, \quad
\tilde{F}=\frac{\alpha_5}{\alpha_1}F,\quad
\tilde{G}=\frac{\alpha_7}{\alpha^2_3}G,
 \end{array}\end{equation} where
$\alpha_l \ (l=1,\dots,8)$ are arbitrary group parameters
($\alpha_{2k-1}\neq0, \ k=1,\dots,4$). \end{consequence}

One sees that (\ref{2-33}) is a nonlinear  system of
functional-differential equations and its general solution seems to
be impossible without further restrictions. For example, assuming a
constant diffusivity $D$, one may derive that $\beta(t,x)$ is linear
with respect to (w.r.t.) the variable $x$; however, $\beta(t,x)$ can
be nonlinear for the  non-constant diffusivity $D$ (see
Tables~\ref{tab-3} and~\ref{tab-4} below for examples).

While the form of the transformations  (\ref{2-30}) is still quite
general, it will be shown in the next section that only  particular
cases play important roles   in solving the  Lie symmetry
classification problem for the nonlinear parabolic-elliptic systems
(\ref{1-3}).

\section{\bf  Lie  symmetries of  a  class of parabolic-elliptic   systems }

 In order to find Lie symmetry operators of   a system  of the form   
(\ref{1a}) via  the Lie  method \cite{arr-15,bl-anco,b-k,fss,olv}, one needs  to  consider manifold $(S_1,S_2)$  
\begin{equation}\label{2-3}  
\begin{array}{l}
 S_1 \equiv   U_t - (D(U)U_{x})_{x}- F(U,V)=\,0, \\  
S_2 \equiv  V_{xx}+G(U,V) =\,0 
\end{array} \end{equation}  
in the space of the following variables:  
$$t, x, U, V, U_t, U_{x}, U_{xx},V_{xx}.   $$

The system  (\ref{1a}) is invariant under the transformations generated by the  
infinitesimal operator  
\begin{equation}\label{2-4}
X=\xi^0 (t, x, U, V)\partial_{t} + \xi^1 (t, x, U, V)\partial_{x} +
\eta^1(t, x, U, V)\partial_{U}+\eta^2(t, x, U, V)\partial_{V}
\end{equation}  
when the following invariance conditions are satisfied:  
\begin{equation}\label{2-5}  
\begin{array}{l}  
\mbox{\raisebox{-1.5ex}{$\stackrel{\displaystyle X}{\scriptstyle
2}$}}\, \left(S_1\right)  
 \equiv  \mbox{\raisebox{-1.5ex}{$\stackrel{\displaystyle X}{\scriptstyle
2}$}}\,  
\left(U_t - (D(U)U_{x})_{x}- F(U,V)\right)  
\Big\vert_{{S_1=0}\atop{S_2=0}}=0, \\[0.3cm]  
\mbox{\raisebox{-1.5ex}{$\stackrel{\displaystyle X}{\scriptstyle
2}$}}\, \left(S_2\right)  
 \equiv  \mbox{\raisebox{-1.5ex}{$\stackrel{\displaystyle X}{\scriptstyle
2}$}}\,  
\left(V_{xx}+G(U,V)\right)  
\Big\vert_{{S_1=0}\atop{S_2=0}}=0.  
\end{array}  
\end{equation}  
The operator $ \mbox{\raisebox{-1.5ex}{$\stackrel{\displaystyle
X}{\scriptstyle
2}$}} $  
is the second  
 prolongation of the operator $X$, i.e.  
\begin{equation}\label{2-6}  
\begin{array} {l}  
\mbox{\raisebox{-1.5ex}{$\stackrel{\displaystyle X}{\scriptstyle
2}$}}  
 = X + \rho_t^1{\partial \over \partial U_{t}}+ \rho_t^1{\partial \over \partial V_{t}}+  
\rho^1_x{\partial \over \partial U_{x}}+ \rho^2_x{\partial \over \partial V_{x}}+\\
\hskip1cm\sigma_{tt}^1\frac{\partial}{\partial
U_{tt}}+\sigma_{tx}^1\frac{\partial}{\partial U_{tx}}+
\sigma_{xx}^1\frac{\partial}{\partial U_{xx}}
+\sigma_{tt}^2\frac{\partial}{\partial
V_{tt}}+\sigma_{tx}^2\frac{\partial}{\partial  V_{tx}}+
\sigma_{xx}^2\frac{\partial}{\partial  V_{xx}},  
\end{array}\end{equation}  
where the coefficients $\rho$ and $\sigma$ with relevant subscripts  
are calculated by well-known formulae (see, e.g., \cite{fss, olv, b-k}).  

Substituting (\ref{2-6})  
 into (\ref{2-5}) and eliminating the derivatives
$U_t$ and $ V_{xx}$ using (\ref{1a}), we can obtain the following
system of determining equations (DEs):

\begin{eqnarray}&& \label{2-7}
 \xi^0_x=\xi^0_U=\xi^0_V=\xi^1_U=\xi^1_V=0,  \\
 && \label{2-8} \eta^1_V=\eta^2_U=\eta^2_{VV}=0,\\
 && \label{2-9} \xi^0_t= 2\xi^1_x- \eta^1\frac{\partial }{\partial U}\ln D,\\
 && \label{2-10} 2\eta^2_{xV}= \xi^1_{xx},\\
 && \label{2-11} 2\eta^1_{xU}+ 2\eta^1_x\frac{\partial }{\partial U}\ln D=
  \xi^1_{xx}- \xi^1_{t}D^{-1},\\
&& \label{2-12} 2\xi^1_{x}- \xi^0_{t}-\eta^1_U
  =\eta^1_{UU}\left(\frac{\partial
}{\partial U}\ln D \right)^{-1} +\eta^1 \frac{\partial }{\partial
U}\ln\frac{\partial D}{\partial
U},\\
&& \label{2-13}\eta^1F_U+\eta^2F_V=\eta^1_t-D\eta^1_{xx}+F\left( \eta^1_U-\xi^0_t\right), \\
&& \label{2-14}\eta^1G_U+\eta^2G_V=-\eta^2_{xx}+G\left(
\eta^2_V-2\xi^1_x\right).
\end{eqnarray}
Obviously, the general solution of above system of DEs essentially
depends on the form of the triplet $(D,F,G)$. Thus, the problem of a
complete description of all possible Lie symmetries (the so called
group classification problem) arises.

First of all we find the so called trivial algebra  (other
terminology used for this algebra is the `principal algebra '  and
the  `kernel of maximal invariance algebras')   of  the class in
question, i.e. the maximal invariance algebra (MAI) admitted by each
equation of the form  (\ref{1a}). Under the natural assumption that
$D, \ F$ and $G$ are arbitrary smooth functions the above system of
DEs can be easily solved and
 the two-dimensional Lie algebra
 generated by the basic operators
     \begin{equation}\label{2-2a}
      X_1 = \partial_t, \ X_2 = \partial_{x}
     \end{equation}
     is obtained.

In order to solve the problem of Lie symmetry classification for
system (\ref{1a}), i.e. to find all possible systems admitting three- and higher-dimensional MAI,
 we use the algorithm based on  form-preserving
transformations, which was briefly discussed in Section 2.
 Moreover, it will be shown below that, similarly to
the case of standard nonlinear reaction-diffusion systems
\cite{ch-king4}, such an approach is  more  efficient also for
parabolic-elliptic systems (compared to with   the classical
Lie--Ovsiannikov algorithm).

Let us formulate a theorem which gives complete information on the  
classical symmetry of the nonlinear system (\ref{1a}).

\begin{theo}\label{th-2}  
All possible  maximal algebras of invariance (up to equivalent
representations generated by transformations of the form
(\ref{2-2})) of the system  (\ref{1a})  for any fixed nonconstant
function $D$
 and nonconstant function vectors $(F, G)$
are presented in  Tables~\ref{tab-1} and~\ref{tab-2}. Any  other system of the form  
(\ref{1a}) with non-trivial Lie symmetry
 is reduced
by a local substitution  of the form
\begin{equation}\label{2-2}  
 \begin{array}  {l} \medskip
 \bar{t}=C_0t + C_1\exp(C_{2}t), \\
\medskip  \bar{x} = C_3x+ C_4\exp(C_5x)+C_6\tan(C_7x), \\
 \bar{U} = C_8+  C_9t+ C_{10}\exp(C_{11}t)U+\\
 \medskip
 \qquad C_4\exp(C_{12}x)U+C_6\cos^3(C_{13}x)U,\\
 \bar{V} = C_{14}+  C_{15}t+ C_{16}\exp(C_{17}t)V+ C_4\exp(C_{12}x)V+ \\
  \qquad C_6\cos^{-1}(C_{13}x)V+C_{18}tx^2+C_{19}\exp(C_{20}t)x^2, \end{array}
    \end{equation}
 to one of those given in Tables~\ref{tab-1} and~\ref{tab-2} (the constants  $C$ with
subscripts are determined by the form of the system in question,
some of them necessarily being zero in any given case). \end{theo}

\textbf{The sketch of the proof.}

Consider the determining equations (\ref{2-7})--(\ref{2-14}) under
the assumption $\frac{\partial D}{\partial U}\neq0$ (we remind the
reader that the case $\frac{\partial D}{\partial U}=0$ was examined
in \cite{ch-king, ch-king1}). Obviously, equations
(\ref{2-7})--(\ref{2-10}) can be easily solved and the formulae
\begin{equation}\label{2-15}  
 \begin{array}  {l} \xi^0=a(t), \ \xi^1=b(t,x), \\
  \eta^1=(2b_x-\dot{a})\left(\frac{\partial}{\partial U}\ln D \right)^{-1}, \ \eta^2=r(t,x)V+p(t,x), \ r_x=\frac{1}{2}\,b_{xx} \end{array}
    \end{equation} are obtained (here
$a, \ b, \ r$ and $p$ are arbitrary  smooth functions and the dot means the time derivative).

Now we can consider equations (\ref{2-11})--(\ref{2-14}) as
classification equations to find the function $D$ and the pairs of
the functions $(F, G)$ for which the system (\ref{1a}) has a
non-trivial Lie symmetry, i.e. its  MAI is larger than   the trivial
algebra (\ref{2-2a}).

It follows from
 (\ref{2-11})--(\ref{2-12}) that there is a special case
$2b_x=a_t$ when the diffusivity
 $D$ can be an arbitrary smooth function.
Indeed equation (\ref{2-11}) produces $b_t=b_{xx}=0$ provided $D$ is
an arbitrary function. Hence, one arrives at  $b=b_1x+b_0, \
a=2b_1t+a_0$, so that the classification equations
(\ref{2-13})--(\ref{2-14}) have the form: \begin{equation}
\label{2-18}
 \begin{array}  {l}  (rV+p)F_V=-2b_1F, \\
  (rV+p)G_V+G\left(2b_1- r\right)=-p_{xx}. \end{array}
    \end{equation}
The general solution of  (\ref{2-18}) depends essentially on the
functions $r$ and  $p$ ($r^2+p^2\neq0$, otherwise a trivial Lie
symmetry is obtained). In order to derive a complete result, one
needs to examine two subcases.

Let us consider the first  subcase $r=0, \ p\neq0,$  i.e. the first equation in (\ref{2-18}) takes the form
$pF_V=-2b_1F$.  A simple analysis says that the general solution of (\ref{2-18}) leads to the functions $F, \ G$ and   coefficients (\ref{2-15}),
 which produce the system and MAI arising in case 1 of Table~\ref{tab-1} provided $b_1\neq0$.
  If $b_1=0$ then case 3 of Table~\ref{tab-2} is derived.

The second subcase $r\neq0$ leads to  case 2 of Table~\ref{tab-1}
and cases 1 and 2 of Table~\ref{tab-2}.

The condition  $2b_x-a_t\neq0$  is a generic inequality and produces
all other cases listed in Tables~\ref{tab-1} and~\ref{tab-2}. Using
this condition and (\ref{2-15}), one easily solves the equation
(\ref{2-12}) and obtains \begin{equation}\label{2-16}
D=\begin{cases} d(U+C_1)^k,
\\  d\exp(C_2U),
   \end{cases}\end{equation}
where $d\neq0, \ C_1, \ C_2$ and $k\neq0$ are arbitrary constants.
Moreover, substituting (\ref{2-16}) into (\ref{2-11}), we obtain
$b=b_1x+b_0, \ b_0, b_1\in \mathbb{R},$ unless
\begin{equation}\label{2-17} D=d(U+C_1)^{-4/3}, \ b=b(x).\end{equation} One
notes that the system (\ref{1a}) with diffusivities (\ref{2-16}) and
(\ref{2-17}) can be simplified to one with $d=1, \ C_1=0, \ C_2=1$
by the appropriate equivalence transformations from (\ref{2-36}).

Now one needs to examine separately other determining equations with power law diffusivity, exponential diffusivity and the special power law coefficient $k=-4/3$. We present here only the last case, i.e. the power low diffusivity $D=U^{-4/3}$, which is the most complicated.

 Let us consider system (\ref{1a}) with
$D=U^{-4/3}.$ It turns out that we may assume that $\xi^1_{xx}\neq0$ (otherwise only particular cases of  (\ref{1a}) with
$D=U^{k}$ will be derived).
The classification equations (\ref{2-13})--(\ref{2-14}) for $D=U^{-4/3}$ take the form
\begin{eqnarray}
&& \label{2-19}
\frac{3}{4}\left(\dot{a}-2b'\right)UF_U+(rV+p)F_V+\left(\frac{\dot{a}}{4}+\frac{3b'}{2}\right)F=
\frac{3}{4}\ddot{a}U+\frac{3b^{(3)}}{2}U^{-1/3}, \\
&&
\label{2-20}\frac{3}{4}\left(\dot{a}-2b'\right)UG_U+(rV+p)G_V+\left(2b'-r\right)G=-\frac{b^{(3)}}{2}V-p_{xx}.
\end{eqnarray}
(because the function $b$ depends only on $x$ we use the standard notations for its derivatives and $b^{(k)}=\frac{d^kb}{dx^k}, k=3,4\dots$).
Now we construct differential consequences of (\ref{2-19})--(\ref{2-20}) w.r.t. $x$:

\begin{eqnarray}
&& \label{2-21}
-\frac{3b''}{2}UF_U+\left(\frac{b''}{2}V+p_x\right)F_V+\frac{3b''}{2}F=
\frac{3b^{(4)}}{2}U^{-1/3}, \\
&&
\label{2-22}-\frac{3b''}{2}UG_U+\left(\frac{b''}{2}V+p_x\right)G_V+\frac{3b''}{2}G=-\frac{b^{(4)}}{2}V-p_{xxx}.
\end{eqnarray}
Taking into account that $b''\neq0$ one easily finds the general
solution of (\ref{2-21})--(\ref{2-22}),
\begin{eqnarray}
&& \label{2-23} F=Uf(\omega)+\frac{3b^{(4)}}{4b''}U^{-1/3}, \ \omega=U\left(V+\frac{2p_x}{b''}\right)^3,\\
&& \label{2-24}
G=Ug(\omega)-\frac{b^{(4)}}{4b''}\left(V+\frac{2p_x}{b''}\right)+\frac{2}{3b''}\left(p_xb^{(4)}-p_{xxx}\right).
\end{eqnarray}
where $f$ and $g$ are arbitrary smooth functions.

The further analysis essentially uses the fact that the functions $F$ and $G$ depend only on $U$ and  $V$
 (not explicitly on $t$ and $x$!). This means that relevant  constraints  on the functions $f, \ g, \ b$  and  $p$ must take place.

Let us consider  the most interesting case when  the functions $f$
and  $ g$ are still arbitrary. This happens if and only if the
constraints
\[  \frac{p_x}{b''}=\alpha_1, \
\frac{b^{(4)}}{b''}=\alpha_2,  \
\frac{2}{3b''}\left(p_xb^{(4)}-p_{xxx}\right)=\alpha_3 \] (here
$\alpha_1, \alpha_2$ and  $\alpha_3$  are some constants) hold in
(\ref{2-23})--(\ref{2-24}). Because the functions
(\ref{2-23})--(\ref{2-24}) must satisfy
 equations
(\ref{2-19})--(\ref{2-20}) for arbitrary functions $f$ and  $ g$,
the additional condition    $p=\gamma r,$  springs up. Direct
calculations show that the constant $\gamma$  arising in the system
obtained  can be reduced to $\gamma=0$ by
equivalence transformation $V+\gamma  \to V$. Thus, we arrive at
system (\ref{1a}) with $D=U^{-4/3}$  and
\begin{eqnarray}
&& \label{2-25} F=Uf(\omega)+\frac{3\alpha}{4}U^{-1/3}, \ \omega=UV^3,\\
&& \label{2-26} G=Ug(\omega)-\frac{\alpha}{4}V,
\end{eqnarray}
i.e. the parabolic-elliptic system
 \begin{equation}\label{2-27}
 \begin{array} {l}
 U_t = (U^{-4/3}U_{x})_{x}+Uf\left(UV^3\right)+\frac{3\alpha}{4}U^{-1/3},\\
0 = V_{xx}+Ug\left(UV^3\right)-\frac{\alpha}{4}V,
 \end{array}\end{equation} where  $f$ and  $ g$ are arbitrary smooth functions, $\alpha$ is an arbitrary constant.
 It turns out that MAI of system (\ref{2-27}) essentially depends on the parameter $\alpha.$

Setting  $\alpha=0$  and substituting (\ref{2-25})--(\ref{2-26})
into (\ref{2-19})--(\ref{2-20}), one obtains the system
\begin{equation}\label{2-28} \dot{a}=0, \ r=\frac{b'}{2}, \ b^{(3)}=0, \end{equation} in
order to find  MAI of (\ref{2-27}). The linear ODE system
(\ref{2-27}) can be easily solved, and as a result case 8 of
Table~\ref{tab-1} is obtained.

 If  $\alpha<0,$  then the equivalence transformation
$\tau=-\frac{\alpha}{4}t, \ y=\frac{\sqrt{-\alpha}}{2}x$ reduces  (\ref{2-27})  to the system listed in case 3
 of Table~\ref{tab-3}.
 If  $\alpha>0,$  then
$\tau=\frac{\alpha}{4}t, \ y=\frac{\sqrt{\alpha}}{2}x$ produces the
system from   case 4 of Table~\ref{tab-3}. It turns out that both
systems
 can be  reduced to one listed in case 8 of Table~\ref{tab-1}.
 The relevant transformations are presented in the 3rd column of
  Table~\ref{tab-3} and they have been derived by applying Theorem~\ref{th-1} to the nonlinear  system (\ref{2-27}).

Thus, the most general  system with the diffusivity $D$ from
(\ref{2-17}) and its MAI  are presented  in case 8 of
Table~\ref{tab-1} and any other system with such a diffusivity
admitting a nontrivial Lie symmetry   is  reduced   to one listed in
case 8 of Table~\ref{tab-1}. It should be noted that the
corresponding substitutions are highly non-trivial and were found
using a special case of form-preserving transformations (not
equivalence transformations!) derived in Theorem~\ref{th-1}.

Now we need to find all possible correctly-specified functions
 $f$ and  $g$ leading to extensions of MAI of system  (\ref{2-27}).
 Such functions have been found by analysis of  the differential consequences  $F_{xV}=F_{tV}=0, \
G_{xU}=G_{tU}=0 $, where the functions $F$ and $G$ are of the form
   (\ref{2-23})--(\ref{2-24}). Finally, we have established only
   two special cases leading to the above mentioned extensions.

{\it(i)} $f=\alpha_1U^{\gamma}V^{3\gamma}, \
g=\alpha_2U^{\gamma+1/3}V^{3\gamma+1}$, where  $\alpha_1, \
\alpha_2$ and $\gamma$ are arbitrary constants
($\alpha^2_1+\alpha^2_2\neq0$). The systems  (\ref{2-27}) with these
 $f$ and  $g$  are  presented in case 9 of Table~\ref{tab-1} and cases  5--6
of Table~\ref{tab-3}.

{\it(ii)} $f=const, \ g=const.$ The constant  $f\neq0$ leads to case
12 of Table~\ref{tab-2} and cases  6--7 of Table~\ref{tab-4}. The
zero constant $f$ leads  to case  13 of Table~\ref{tab-2} and cases
8--9 of Table~\ref{tab-4}.

At the final stage, we note that the form-preserving transformations
derived above are still working, so that  only case 9 of
Table~\ref{tab-1} and cases  12--13 of Table~\ref{tab-2} are
essentially new.

Substituting all possible functions $D$ from (\ref{2-16}) (with
$d=1, \ C_1=0, \ C_2=1$) into (\ref{2-13})--(\ref{2-14}) and taking
into account (\ref{2-15}), we construct the corresponding pairs $(F,
G)$ and Lie algebras. The results are summarised in the form of
cases 3--7, 10--14 in Table~\ref{tab-1} and cases 4--11, 14--21 in
Table~\ref{tab-2}. All other parabolic-elliptic   systems of the
current class with non-trivial Lie symmetry can be reduced to those
in Tables~\ref{tab-1}--\ref{tab-2} by the appropriate equivalence
transformations from (\ref{2-36}) and/or form-preserving
transformations and these are listed in
Tables~\ref{tab-3}--\ref{tab-4}.

The sketch of the proof is now completed. \ $\Box$

\begin{table}
\caption{ Lie symmetries of system (\ref{1a}) in the case
$F_VG_U\neq0$} \label{tab-1}
\begin{center}
\begin{tabular}{|c|c|c| }  \hline
 Case & RD system & Basic operators of MAI\\
\hline
   1. &$U_t=\left(D(U)U_x\right)_x+e^Vf(U)$&
 $\partial_t, \ \partial_x, \ 2t\partial_t+x\partial_x-2\partial_V$\\&$0=V_{xx}+e^Vg(U)$&\\
  \hline
2. &$U_t=\left(D(U)U_x\right)_x+V^\beta f(U) $&
 $\partial_t, \ \partial_x, \ 2\beta t\partial_t+\beta x\partial_x-2V\partial_V$\\&$0=V_{xx}+V^{\beta+1}g(U)$&\\
  \hline
  3. &$U_t=\left(U^kU_x\right)_x+U^{\gamma+1} f\left(e^VU^\alpha\right) $&
 $\partial_t, \ \partial_x,$ \\&$0=V_{xx}+U^{\gamma-k} g\left(e^VU^\alpha\right)$&$2\gamma t\partial_t+(\gamma-k) x\partial_x-2U\partial_U+2\alpha \partial_V$\\
  \hline
  4. &$U_t=\left(U^kU_x\right)_x+U^{\gamma+1} f\left(VU^\beta\right) $&
 $\partial_t, \ \partial_x, $\\&$0=V_{xx}+VU^{\gamma-k} g\left(VU^\beta\right)$&$2\gamma t\partial_t+(\gamma-k) x\partial_x-2U\partial_U+2\beta V\partial_V$\\
  \hline
   5. &$U_t=\left(U^kU_x\right)_x+\alpha_1e^VU^{\gamma+1} $&
 $\partial_t, \ \partial_x,  \ 2t\partial_t+x\partial_x-2\partial_V,$\\&$0=V_{xx}+\alpha_2e^VU^{\gamma-k}$&$kt\partial_t-U\partial_U+(\gamma-k) \partial_V$\\
  \hline
   6. &$U_t=\left(U^kU_x\right)_x+\alpha_1V^\beta U^{\gamma+1} $&
 $\partial_t, \ \partial_x,  \ 2\beta t\partial_t+\beta x\partial_x-2V\partial_V,$\\&$0=V_{xx}+\alpha_2V^{\beta+1}U^{\gamma-k}$&$k\beta t\partial_t-\beta U\partial_U+(\gamma-k)V\partial_V$\\
  \hline
   7. &$U_t=\left(U^kU_x\right)_x+\alpha_1 U^{k+1}+UV $&
 $\partial_t, \ \partial_x,  \ -k\varphi(t)\partial_t+\varphi'(t)U\partial_U+$\\&$0=V_{xx}+
 \alpha_2U^{k}$&$\left(k\varphi'(t)V+(k+1)\varphi''(t)\right)\partial_V$\\
  \hline
  8. &$ \ U_t=\left(U^{-4/3}U_x\right)_x+U f(UV^3)$&
 $\partial_t, \ \partial_x, \ 2x\partial_x-3U\partial_U+ V\partial_V,
 $\\&$0=V_{xx}+ U g(UV^3)$&$x^2\partial_x-3xU\partial_U+ xV\partial_V$\\
  \hline
   9. &$ \ U_t=\left(U^{-4/3}U_x\right)_x+\alpha_1U^{1+\gamma}V^{3\gamma}$&
 $\partial_t, \ \partial_x, \ 2x\partial_x-3U\partial_U+ V\partial_V,
 $\\&$0=V_{xx}+ \alpha_2U^{4/3+\gamma}V^{3\gamma+1}$&$x^2\partial_x-3xU\partial_U+ xV\partial_V,$
 \\ &&
 $4\gamma t\partial_t+3\gamma U\partial_U- (4/3+\gamma)V\partial_V $\\
  \hline
  10. &$U_t=\left(e^UU_x\right)_x+e^{(\gamma+1)U}
f\left(V+\alpha U\right) $&
 $\partial_t, \ \partial_x, $\\&$0=V_{xx}+e^{\gamma U}
g\left(V+\alpha U\right)$&$2(\gamma+1) t\partial_t+\gamma x\partial_x-2\partial_U+2\alpha \partial_V$\\
  \hline
   11. &$U_t=\left(e^UU_x\right)_x+e^{(\gamma+1)U}f\left(Ve^{\beta U}\right) $&
 $\partial_t, \ \partial_x, $\\&$0=V_{xx}+Ve^{\gamma U} g\left(Ve^{\beta U}\right)$&$2(\gamma+1) t\partial_t+\gamma x\partial_x-2\partial_U+2\beta V\partial_V$\\
  \hline
  12. &$U_t=\left(e^UU_x\right)_x+\alpha_1\exp\left((\gamma+1)U+V\right) $&
 $\partial_t, \ \partial_x,  \ 2t\partial_t+x\partial_x-2\partial_V,$\\&$0=V_{xx}+\alpha_2\exp\left(\gamma U+V\right)$&$t\partial_t-\partial_U+\gamma \partial_V$\\
  \hline
  13. &$U_t=\left(e^UU_x\right)_x+\alpha_1e^{(\gamma+1)U}V^\beta $&
 $\partial_t, \ \partial_x,  \ 2\beta t\partial_t+\beta x\partial_x-2V\partial_V,$\\&$0=V_{xx}+\alpha_2e^{\gamma U}V^{\beta+1}$&$\beta t\partial_t-
 \beta \partial_U+\gamma V\partial_V$\\
  \hline
  14. &$U_t=\left(e^UU_x\right)_x+\alpha_1 e^U+V $&
 $\partial_x,  \ \varphi(t)\partial_t-\varphi'(t)\partial_U-$\\&$0=V_{xx}+
 \alpha_2e^U$&$\left(\varphi'(t)V+\varphi''(t)\right)\partial_V$\\
  \hline
      \end{tabular}
\end{center}
\end{table}

\begin{remark}\label{rem-1} Each case in Table~\ref{tab-1} contains  semi-coupled systems
(i.e. those with $F_VG_U=0$) as special subcases.  A system
specified in such a way has the same MAI provided it is not listed
in Table~\ref{tab-2} as a special case. For example, the system
arising in case 1 with an arbitrary function $f(U)$  and
$g(U)=constant$ is a semi-coupled system; however, this is not
listed in Table~\ref{tab-2}, so that its MAI is still a
3-dimensional Lie algebra. \end{remark}

\begin{table}
\caption{  Lie symmetries of system (\ref{1a}) in the case
$F_VG_U=0$ and  $\left(F_V\right)^2+\left(G_U\right)^2\neq0$}
\label{tab-2}
\begin{center}
\begin{tabular}{|c|c|c| }  \hline
 Case & RD system & Basic operators of MAI\\
\hline 1. &$U_t=\left(D(U)U_x\right)_x$&
 $\partial_t, \ \partial_x, \ 2t\partial_t+x\partial_x+2V\partial_V, $\\&$0=V_{xx}+g(U)$&$\varphi(t)x\partial_V, \ \psi(t)\partial_V$\\
  \hline
 2. &$U_t=\left(D(U)U_x\right)_x+f(U)$&
 $\partial_t, \ \partial_x, \ \psi(t)V\partial_V$\\&$0=V_{xx}+Vg(U)$&\\
  \hline
   3. &$U_t=\left(D(U)U_x\right)_x+f(U)$&
 $\partial_t, \ \partial_x, \ h(t,x)\partial_V, \ h_{xx}+\alpha h=0$\\&$0=V_{xx}+\alpha V+g(U)$&\\
  \hline 4. &$U_t=\left(U^kU_x\right)_x+\alpha_0U^{\beta+1} $&
 $\partial_t, \ \partial_x,  \ 2\beta t\partial_t+(\beta-k) x\partial_x-2U\partial_U,$\\&$0=V_{xx}+\alpha VU^{\beta-k}, \ \alpha\neq0$&$\psi(t)V\partial_V$\\
  \hline
  5. &$U_t=\left(U^kU_x\right)_x+UV $&
 $\partial_x,  \   kx\partial_x+2U\partial_U, \ -k\varphi(t)\partial_t+ $\\&$0=V_{xx}$&$\varphi'(t)U\partial_U+ (k\varphi'(t)V+(k+1)\varphi''(t))\partial_V$\\
  \hline
   6. &$U_t=\left(U^kU_x\right)_x+\alpha_1 U^{k+1}$&
 $\partial_t, \ \partial_x,  \
 -kt\partial_t+U\partial_U+h^0(t,x)\partial_V, $\\&$0=V_{xx}+\ln U+
 \alpha_2V$&$ h(t,x)\partial_V$\\
  \hline
 7. &$U_t=\left(U^kU_x\right)_x+\alpha_1 U^{k+1}$&
 $\partial_t, \ \partial_x,  \
 -kt\partial_t+U\partial_U+\beta V\partial_V, $\\&$0=V_{xx}+ U^\beta+
 \alpha_2V$&$h(t,x)\partial_V$\\
  \hline
  8. &$U_t=\left(U^kU_x\right)_x+\alpha U^{\gamma+1}, \
\alpha\gamma\neq0$&
 $\partial_t, \ \partial_x,  \
 2\gamma t\partial_t+(\gamma-k)x\partial_x-2U\partial_U+$\\&$0=V_{xx}+ \ln U$&$\left(2(\gamma-k)V+x^2\right) \partial_V, \
 \varphi(t)x\partial_V, \ \psi(t)\partial_V$\\
  \hline
   9. &$U_t=\left(U^kU_x\right)_x+\alpha U^{\gamma+1}, \
\alpha\gamma\neq0$&
 $\partial_t, \ \partial_x,  \
 2\gamma t\partial_t+(\gamma-k)x\partial_x-2U\partial_U+$\\&$0=V_{xx}+ U^\beta$&$2(\gamma-\beta-k)V\partial_V, \
 \varphi(t)x\partial_V, \ \psi(t)\partial_V$\\
  \hline
   10. &$ \ U_t=\left(U^kU_x\right)_x$&
 $\partial_t, \ \partial_x,  \
 -2k t\partial_t+2U\partial_U-x^2\partial_V,$\\&$0=V_{xx}+ \ln U$&$2t\partial_t+x\partial_x+2V\partial_V, \
 \varphi(t)x\partial_V, \ \psi(t)\partial_V$\\
  \hline
   11. &$ \ U_t=\left(U^kU_x\right)_x$&
 $\partial_t, \ \partial_x,  \ -k t\partial_t+U\partial_U+\beta V\partial_V,
 $\\&$0=V_{xx}+ U^\beta$&$2t\partial_t+x\partial_x+2V\partial_V, \
 \varphi(t)x\partial_V, \ \psi(t)\partial_V$\\
  \hline
    12. &$ \ U_t=\left(U^{-4/3}U_x\right)_x+\alpha U, \ \alpha\neq0$&
 $\partial_t, \ \partial_x, \ 2x\partial_x-3U\partial_U+ V\partial_V,
 $\\&$0=V_{xx}+ U$&$x^2\partial_x-3xU\partial_U+ xV\partial_V,$
 \\ &&
 $\varphi(t)x\partial_V, \ \psi(t)\partial_V$\\
  \hline
  13. &$ \ U_t=\left(U^{-4/3}U_x\right)_x$&
 $\partial_t, \ \partial_x, \ 2x\partial_x-3U\partial_U+ V\partial_V,
 $\\&$0=V_{xx}+ U$&$x^2\partial_x-3xU\partial_U+ xV\partial_V,$
 \\ &&
 $4t\partial_t+3\left(U\partial_U+V\partial_V\right), \ \varphi(t)x\partial_V, \ \psi(t)\partial_V$\\
  \hline
  14.
&$U_t=\left(e^UU_x\right)_x+\alpha_0e^{(\beta+1)U} $&
 $\partial_t, \ \partial_x,  \ 2(\beta+1) t\partial_t+\beta x\partial_x-2\partial_U,$\\&$0=V_{xx}+\alpha Ve^{\beta U}, \ \alpha\neq0$&$\psi(t)V\partial_V$\\
  \hline
  15. &$U_t=\left(e^UU_x\right)_x+V $&
 $\partial_t, \ \partial_x,  \   x\partial_x+2\partial_U, $\\&$0=V_{xx}$&
 $\varphi(t)\partial_t-\varphi'(t)\partial_U-\left(\varphi'(t)V+\varphi''(t)\right)\partial_V$\\
  \hline
   16. &$U_t=\left(e^UU_x\right)_x+\alpha_1 e^U$&
 $\partial_t, \ \partial_x,  \
 -t\partial_t+\partial_U+h^0(t,x)\partial_V, \ h(t,x)\partial_V$\\&$0=V_{xx}+U+
 \alpha_2V$&$$\\
  \hline \end{tabular}
\end{center}
\end{table} \begin{table}
\leftline{Continuation of Table~\ref{tab-2}} \medskip
\begin{center}
\begin{tabular}{|c|c|c| }  \hline 17. &$U_t=\left(e^UU_x\right)_x+\alpha_1 e^U$&
 $\partial_t, \ \partial_x,  \
 -t\partial_t+\partial_U+\beta V\partial_V, \ h(t,x)\partial_V$\\&$0=V_{xx}+ e^{\beta U}+
 \alpha_2V$&$$\\
  \hline
 18. &$U_t=\left(e^UU_x\right)_x+\alpha e^{\gamma U}, \
\alpha\gamma\neq0$&
 $\partial_t, \ \partial_x,  \
 2\gamma t\partial_t+(\gamma-1)x\partial_x-2\partial_U+$\\&$0=V_{xx}+ U$&$\left(2(\gamma-1)V+x^2\right) \partial_V, \
 \varphi(t)x\partial_V, \ \psi(t)\partial_V$\\
  \hline
  19. &$U_t=\left(e^UU_x\right)_x+\alpha e^{\gamma U}, \
\alpha\gamma\neq0$&
 $\partial_t, \ \partial_x,  \
 2\gamma t\partial_t+(\gamma-1)x\partial_x-2\partial_U+$\\&$0=V_{xx}+ e^{\beta U}$&$2(\gamma-\beta-1)V\partial_V, \
 \varphi(t)x\partial_V, \ \psi(t)\partial_V$\\
  \hline
   20. &$ \ U_t=\left(e^UU_x\right)_x$&
 $\partial_t, \ \partial_x,  \
 2 t\partial_t-2\partial_U+x^2\partial_V,$\\&$0=V_{xx}+ U$&$2t\partial_t+x\partial_x+2V\partial_V, \
 \varphi(t)x\partial_V, \ \psi(t)\partial_V$\\
  \hline
  21. &$ \ U_t=\left(e^UU_x\right)_x$&
 $\partial_t, \ \partial_x,  \  t\partial_t-\partial_U-\beta V\partial_V,
 $\\&$0=V_{xx}+ e^{\beta U}$&$2t\partial_t+x\partial_x-\frac{2}{\beta}\partial_U, \
 \varphi(t)x\partial_V, \ \psi(t)\partial_V$\\
  \hline
\end{tabular}
\end{center}
\end{table}

\begin{remark}\label{rem-2} In Tables~\ref{tab-1} and~\ref{tab-2}, $\alpha,  \alpha_0, \alpha_1,
\alpha_2, \beta, \gamma, k $ are arbitrary constants
($k\beta\left((\alpha_1)^2+(\alpha_2)^2\right)\neq0$); $D, f, g,
\varphi $ and $\psi$ are arbitrary smooth functions of the relevant
arguments;
 the function $h(t,x)$ is  an arbitrary solution of equation
$h_{xx}+\alpha_2h=0,$ while the function $h^{0}(t,x)$ is an
arbitrary solution of the equation $h^{0}_{xx}+\alpha_2h^{0}+1=0$.
\end{remark}
 \begin{table}
\caption{RD systems  that are reduced to those in Table~\ref{tab-1}
 by the form-preserving transformations } \label{tab-3}
\begin{center}
\begin{tabular}{|c|c|c|c| }  \hline &
 RD system & Transformation of variables & Case of  Table~\ref{tab-1} \\
\hline 1.&
$u_{\tau}=\left(u^ku_y\right)_y+u^{k+1}f\left(e^vu^\alpha\right)+\lambda
u$ &$U=e^{-\lambda \tau}u, \ V=v+\alpha\lambda \tau,$&
 3   \\ & $0=v_{yy}+g\left(e^vu^\alpha\right)$&$t=\frac{1}{k\lambda}\,e^{k\lambda \tau}$& with $\gamma=k$\\
  \hline 2. &
 $u_{\tau}=\left(u^ku_y\right)_y+u^{k+1}f\left(vu^\beta\right)+\lambda u$
&$U=e^{-\lambda \tau}u, \ V=e^{\beta\lambda \tau}v,$&
 4   \\ & $0=v_{yy}+vg\left(vu^\alpha\right)$&$t=\frac{1}{k\lambda}\,e^{k\lambda \tau}$& with $\gamma=k$\\
  \hline 3. &
   $u_{\tau}=\left(u^{-4/3}u_y\right)_y+uf\left(uv^3\right)-3u^{-1/3}$
&$U=\cos^3y\,u, \ V=\cos^{-1}y\,v,$&
 8  \\ & $0=v_{yy}+ug\left(uv^3\right)+v$&$x=\tan y$&\\
  \hline 4. &
    $u_{\tau}=\left(u^{-4/3}u_y\right)_y+uf\left(uv^3\right)+3u^{-1/3}$
&$U=e^{3y}u, \ V=e^{-y}v,$&
 8  \\ & $0=v_{yy}+ug\left(uv^3\right)-v$&$x=\frac{1}{2}\,e^{-2y}$&\\
  \hline 5. &
    $u_{\tau}=\left(u^{-4/3}u_y\right)_y+\alpha_1u^{1+\gamma}v^{3\gamma}-3u^{-1/3}$
&$U=\cos^3y\,u, \ V=\cos^{-1}y\,v,$&
 9  \\ &$0=v_{yy}+\alpha_2u^{4/3+\gamma}v^{3\gamma+1}+v$&$x=\tan y$&\\
  \hline 6. &
      $u_{\tau}=\left(u^{-4/3}u_y\right)_y+\alpha_1u^{1+\gamma}v^{3\gamma}+3u^{-1/3}$
&$U=e^{3y}u, \ V=e^{-y}v,$&
 9  \\ & $0=v_{yy}+\alpha_2u^{4/3+\gamma}v^{3\gamma+1}-v$&$x=\frac{1}{2}\,e^{-2y}$&\\
  \hline 7. &
  $u_{\tau}=\left(e^uu_y\right)_y+e^{u}f\left(v+\alpha u\right)+\lambda$
&$U=u-\lambda \tau, \ V=v+\alpha\lambda \tau,$&
 10   \\ & $0=v_{yy}+g\left(v+\alpha u\right)$&$t=\frac{1}{\lambda}\,e^{\lambda \tau}$& with $\gamma=0$\\
  \hline 8. &
  $u_{\tau}=\left(e^uu_y\right)_y+e^{u}f\left(ve^{\beta u}\right)+\lambda$
&$U=u-\lambda \tau, \ V=e^{\beta\lambda \tau}v,$&
 11   \\ & $0=v_{yy}+vg\left(ve^{\beta u}\right)$&$t=\frac{1}{\lambda}\,e^{\lambda \tau}$& with $\gamma=0$\\
  \hline
                    \end{tabular}
\end{center}
\end{table}
\begin{table}
\caption{RD systems  that are reduced to those in Table~\ref{tab-2}
 by the form-preserving transformations} \label{tab-4}
\begin{center}
\begin{tabular}{|c|c|c|c| }  \hline &
 RD system & Transformation of variables & Case of  Table~\ref{tab-2} \\
 \hline 1. &
   $u_{\tau}=\left(u^ku_y\right)_y+\alpha_1u^{k+1}+\lambda u$
&$U=e^{-\lambda \tau}u, \ t=\frac{1}{k\lambda}\,e^{k\lambda \tau},$&
 6   \\ & $0=v_{yy}+\ln u+\alpha_2v$&$V=
\begin{cases}v+\frac{\lambda}{\alpha_2}\,\tau, & $if$ \ \alpha_2\neq0, \\  v+\frac{\lambda}{2}\,\tau y^2,
  & $if$ \  \alpha_2=0  \end{cases}$&\\
  \hline 2. &
  $u_{\tau}=\left(u^ku_y\right)_y+\alpha_1u^{k+1}+\lambda u$
&$U=e^{-\lambda \tau}u, \ t=\frac{1}{k\lambda}\,e^{k\lambda \tau},$&
 7   \\ & $0=v_{yy}+ u^\beta+\alpha_2v+\gamma$&$V=
\begin{cases}\left(v+\frac{\gamma}{\alpha_2}\right)e^{-\beta\lambda \tau}, & $if$ \ \alpha_2\neq0, \\
\left(v+\frac{\gamma}{2}\,y^2\right)e^{-\beta\lambda \tau},
  & $if$ \  \alpha_2=0  \end{cases}$&\\
  \hline 3. &
  $u_{\tau}=\left(u^ku_y\right)_y+\alpha u^{\gamma+1}$ &$V=
v+\frac{\lambda}{2}\,y^2$& 9, \ if $\alpha\neq0$
\\ & $0=v_{yy}+ u^\beta+\lambda$&&11, \ if $\alpha=0$\\
  \hline 4. &
   $u_{\tau}=\left(u^ku_y\right)_y+\lambda u$
&$U=e^{-\lambda \tau}u, \ t=\frac{1}{k\lambda}\,e^{k\lambda \tau},$&
 10   \\ & $0=v_{yy}+\ln u$&$V=
v+\frac{\lambda}{2}\,\tau y^2$&\\
  \hline 5. &
   $u_{\tau}=\left(u^ku_y\right)_y+\lambda u$
&$U=e^{-\lambda \tau}u, \ t=\frac{1}{k\lambda}\,e^{k\lambda \tau},$&
 11   \\ & $0=v_{yy}+ u^\beta+\gamma$&$V=
\left(v+\frac{\gamma}{2}\,y^2\right)e^{-\beta\lambda \tau}$&\\
  \hline 6. &
   $u_{\tau}=\left(u^{-4/3}u_y\right)_y+\alpha u-3u^{-1/3}$
&$U=\cos^3y\,u, \ V=\cos^{-1}y\,v,$&
 12  \\ & $0=v_{yy}+u+v$&$x=\tan y$&\\
  \hline 7. &
          $u_{\tau}=\left(u^{-4/3}u_y\right)_y+\alpha u+3u^{-1/3}$
&$U=e^{3y}u, \ V=e^{-y}v,$&
 12  \\ & $0=v_{yy}+u-v$&$x=\frac{1}{2}\,e^{-2y}$&\\
  \hline 8.  &
          $u_{\tau}=\left(u^{-4/3}u_y\right)_y-3u^{-1/3}$
&$U=\cos^3y\,u, \ V=\cos^{-1}y\,v,$&
 13  \\ & $0=v_{yy}+u+v$&$x=\tan y$&\\
  \hline 9. &
            $u_{\tau}=\left(u^{-4/3}u_y\right)_y+3u^{-1/3}$
&$U=e^{3y}u, \ V=e^{-y}v,$&
 13  \\ & $0=v_{yy}+u-v$&$x=\frac{1}{2}\,e^{-2y}$&\\
  \hline 10. &
   $u_{\tau}=\left(e^uu_y\right)_y+\alpha_1e^{u}+\lambda$
&$U=u-\lambda \tau, \ t=\frac{1}{\lambda}\,e^{\lambda \tau},$&
 16  \\ & $0=v_{yy}+u+\alpha_2v$&$V=
\begin{cases}v+\frac{\lambda}{\alpha_2}\,\tau, & $if$ \ \alpha_2\neq0, \\  v+\frac{\lambda}{2}\,\tau y^2,
  & $if$ \  \alpha_2=0  \end{cases}$&\\
  \hline 11. &
    $u_{\tau}=\left(e^uu_y\right)_y+\alpha_1e^{u}+\lambda$
&$U=u-\lambda \tau, \ t=\frac{1}{\lambda}\,e^{\lambda \tau},$&
 17  \\ & $0=v_{yy}+e^{\beta u}+\alpha_2v+\gamma$&$V=
\begin{cases}\left(v+\frac{\gamma}{\alpha_2}\right)e^{-\beta\lambda \tau}, & $if$ \ \alpha_2\neq0, \\
\left(v+\frac{\gamma}{2}\,y^2\right)e^{-\beta\lambda \tau},
  & $if$ \  \alpha_2=0  \end{cases}$&\\
  \hline 12. &
  $u_{\tau}=\left(e^uu_y\right)_y+\alpha e^{\gamma u}$ &$V=
v+\frac{\lambda}{2}\,y^2$& 19, \ if $\alpha\neq0$ \\ & $0=v_{yy}+
e^{\beta u}+\lambda$&&21, \ if $\alpha=0$
  \\
  \hline 13. &
   $u_{\tau}=\left(e^uu_y\right)_y+\lambda$
&$U=u-\lambda \tau, \ t=\frac{1}{\lambda}\,e^{\lambda \tau},$&
 20  \\ & $0=v_{yy}+u$&$V=
v+\frac{\lambda}{2}\,\tau y^2$&\\
  \hline 14. &
    $u_{\tau}=\left(e^uu_y\right)_y+\lambda$ &$U=u-\lambda \tau, \ t=\frac{1}{\lambda}\,e^{\lambda \tau},$&
 21  \\ & $0=v_{yy}+e^{\beta u}+\gamma$&$V=
\left(v+\frac{\gamma}{2}\,y^2\right)e^{-\beta\lambda \tau}$&\\
  \hline
 \end{tabular}
\end{center}
\end{table}
\begin{remark}\label{rem-3} The systems presented in the 2nd column of Tables~\ref{tab-3}
 and~\ref{tab-4} are presented in their  simplified  forms taking into
account the equivalence transformations (\ref{2-36}). \end{remark}

We would like to conclude this section by the following observation.
As it was noted in Section 2,
 the form-preserving transformations are used in order to solve the problem of the
 Lie symmetry classification for
the class of systems (\ref{1a}). We have proved that there are 35
inequivalent systems admitting non-trivial Lie algebras (i.e. their
MAI is three- and higher-dimensional) and they are listed in
Tables~\ref{tab-1} and~\ref{tab-2}. The systems  are inequivalent up
to point transformations of the form (\ref{2-2}), which are
particular cases of form-preserving transformations described in
Theorem~\ref{th-1}. It is a standard routine to show that there are
no  other point transformations that allow the reduction of a system
from Table~\ref{tab-1} or~\ref{tab-2} to another system from these
tables. Thus, a canonical list of nonlinear parabolic-elliptic
systems of the form (\ref{1a}) admitting a non-trivial Lie symmetry
consists of the 35 inequivalent systems listed in Tables~\ref{tab-1}
and~\ref{tab-2}.

A natural question arises: How many systems of the form (\ref{1a}) with non-trivial
 Lie symmetry can be derived using the Lie--Ovsiannikov algorithm?
 This means that we should take into account only the group of equivalence transformations (\ref{2-36}).
A formal  answer is very simple: one should   extend the list of
above 35 systems by those from Tables~\ref{tab-3} and~\ref{tab-4},
therefore 57 systems with the three- and higher-dimensional MAI will
be derived (in the sense that
the relevant MAI are inequivalent up to the equivalence transformations (\ref{2-36})).

\section{Boundary value problems for a one-dimensional tumour growth model with negligible cell viscosity}

In this section we apply the results of Lie symmetry classification
derived in Section 2 for exact solving a real world model proposed
in  \cite{by-ki-2003} (see p. 570).

Consider the pressure difference with $r=1$  \cite{by-ki-2003} (see
p. 569 therein)
\[ \Sigma(\alpha) = k_0 \frac {\alpha - \alpha_* } {(1- \alpha )^q} \]
assuming $ \alpha > \alpha_*$. Setting
\[ k(\alpha) = k_0\alpha^{1-m}(1- \alpha )^2 \frac{d (\alpha \Sigma)}{d\alpha} \]
the equation (see (23) in \cite{by-ki-2003}) for the tumour cell
concentration, $\alpha(t,x)$,  takes the form
 \begin{equation}\label{0}
 \alpha_t = \left(\alpha^m \alpha_{x}\right)_{x}+ S(\alpha, c).
 \end{equation}

In order to simplify the boundary conditions we set
 \begin{equation}\label{*}
 U=\alpha - \alpha_*,  \quad  V=  c_{\infty} - c
 \end{equation}
(wherein $c$ corresponds to the level of nutrient in
\cite{by-ki-2003}, but might instead correspond to that of a
chemotherapeutic  drug) and  arrive at the BVP
 \begin{equation}\label{4-1a}
 \begin{array} {l}
 U_t = \left((U+\alpha_*)^mU_{x}\right)_{x}+S(U+\alpha_*,c_{\infty}-V),\\
0 = V_{xx} +Q(U+\alpha_*,c_{\infty}-V),\\
x=0: \ U_x=V_x=0,\\
x=R(t): \ U=V=0,\\
x=R(t): \ R'=-\alpha^{m-1}_*U_x,
 \end{array}\end{equation}
 where the concentrations $U(t,x)$ and $V(t,x)$  and the moving boundary location $R(t)$ are unknown functions, while the functions $S$ and $Q$ are given smooth functions and $m$,
 $\alpha_*$ and $ c_{\infty}$ are nonnegative parameters.

 One may note that the governing equations of this
   BVP with $S=f(U)V^{-\beta}$ and   $Q=g(U)V^{-\beta+1}$ are  a
    particular case  of case 2 in Table~\ref{tab-1}, so that they admit  %
the  Lie symmetry operator
 \begin{equation}\label{1c} X=2\beta t\partial_t+\beta x\partial_x + 2V\partial_V.\end{equation}
 It turns out that the following statement can be easily proved,
  using the definition of invariance for BVPs with moving boundaries \cite{ch-kov11a}.

  \begin{theo}\label{th-3}
  The nonlinear  BVP (\ref{4-1a}) with $S=f(U)V^{-\beta}$ and   $Q=g(U)V^{-\beta+1}$  is invariant with respect to the two-dimensional MAI generated by the Lie symmetry operators  (\ref{1c}) and $\partial_t$.
 \end{theo}

  Clearly the operator $\partial_t$ cannot help to find any
   realistic solutions of  BVP (\ref{4-1a}), while the
    Lie symmetry  (\ref{1c}) guarantees
    a reduction of the problem, which leads to some  interesting results.
 In fact, the operator generates the ansatz
 \begin{equation}\label{1b}
 \begin{array} {l}
 U = \varphi(\omega), \ \omega=\frac{x}{\sqrt{t}}\,,\\
V = t^{\frac{1}{\beta}}\psi(\omega),
 \end{array}\end{equation}
which immediately specifies the function
 $R(t)=\omega_0\sqrt{t}$\,  with  the constant  $\omega_0>0$ to be found.
 The reduced ODE problem is
 \begin{equation}\label{2}
 \begin{array} {l}
(\varphi+\alpha_*)^m\varphi''+m(\varphi+\alpha_*)^{m-1}\left(\varphi'\right)^2+\frac{\omega}{2}\,\varphi'+f(\varphi)
\psi^{-\beta}=0,\\
\psi''+g(\varphi)\psi^{-\beta+1}=0,\\
\omega=0: \ \varphi'=\psi'=0,\\
\omega=\omega_0: \ \varphi=\psi=0, \
\varphi'=-\frac{\alpha^{1-m}_*\omega_0}{2}.
 \end{array}\end{equation}

The nonlinear BVP (\ref{2}) is still  a complicated problem and we
were unable to solve it analytically in general. Let us consider a
special case, namely  (\ref{2})  with  $\beta=1:$
 \begin{equation}\label{2a}
 \begin{array} {l}
(\varphi+\alpha_*)^m\varphi''+m(\varphi+\alpha_*)^{m-1}\left(\varphi'\right)^2+
\frac{\omega}{2}\,\varphi'+f(\varphi)\psi^{-1}=0,\\
\psi''+g(\varphi)=0,\\
\omega=0: \ \varphi'=\psi'=0,\\
\omega=\omega_0: \ \varphi=\psi=0, \
\varphi'=\frac{\alpha^{1-m}_*\omega_0}{2}.
 \end{array}\end{equation}

In this case the exact solution can be found for  arbitrary $m$ in
the form
 \begin{equation}\label{3}
 \begin{array} {l}
 \varphi=\frac{\alpha^{1-m}_*}{4}\left(\omega^2_0-\omega^2\right),  \ \omega< \omega_0,\\
\psi= \frac{\alpha^{1-m}_*
q_0}{48}\left(\omega^2_0-\omega^2\right)\left(5\omega^2_0-\omega^2\right),
\ \omega< \omega_0
 \end{array}\end{equation}
provided the functions $ f(\varphi)$  and  $g(\varphi)$  have the form
 \begin{equation}\label{4}
 \begin{array} {l}
 g= q_0\varphi,  \quad  q_0>0, \\
f= \frac{q_0\alpha^{m-1}_*}{3}\varphi\left(\varphi+\alpha^{1-m}_*\omega^2_0\right)\\
\quad  \times \Bigg(
(m+\frac{1}{2})\alpha^{1-m}_*(\varphi+\alpha_*)^m-
\frac{m\alpha^{2-m}_*}{4}\left(\alpha^{-m}_*\omega^2_0+4\right)(\varphi+\alpha_*)^{m-1}-
\varphi+\frac{\alpha^{1-m}_*}{4}\omega^2_0\Bigg).
 \end{array}\end{equation}

\begin{figure}
\begin{minipage}{7cm}
  \quad  \quad \begin{center}\includegraphics[width=8cm]{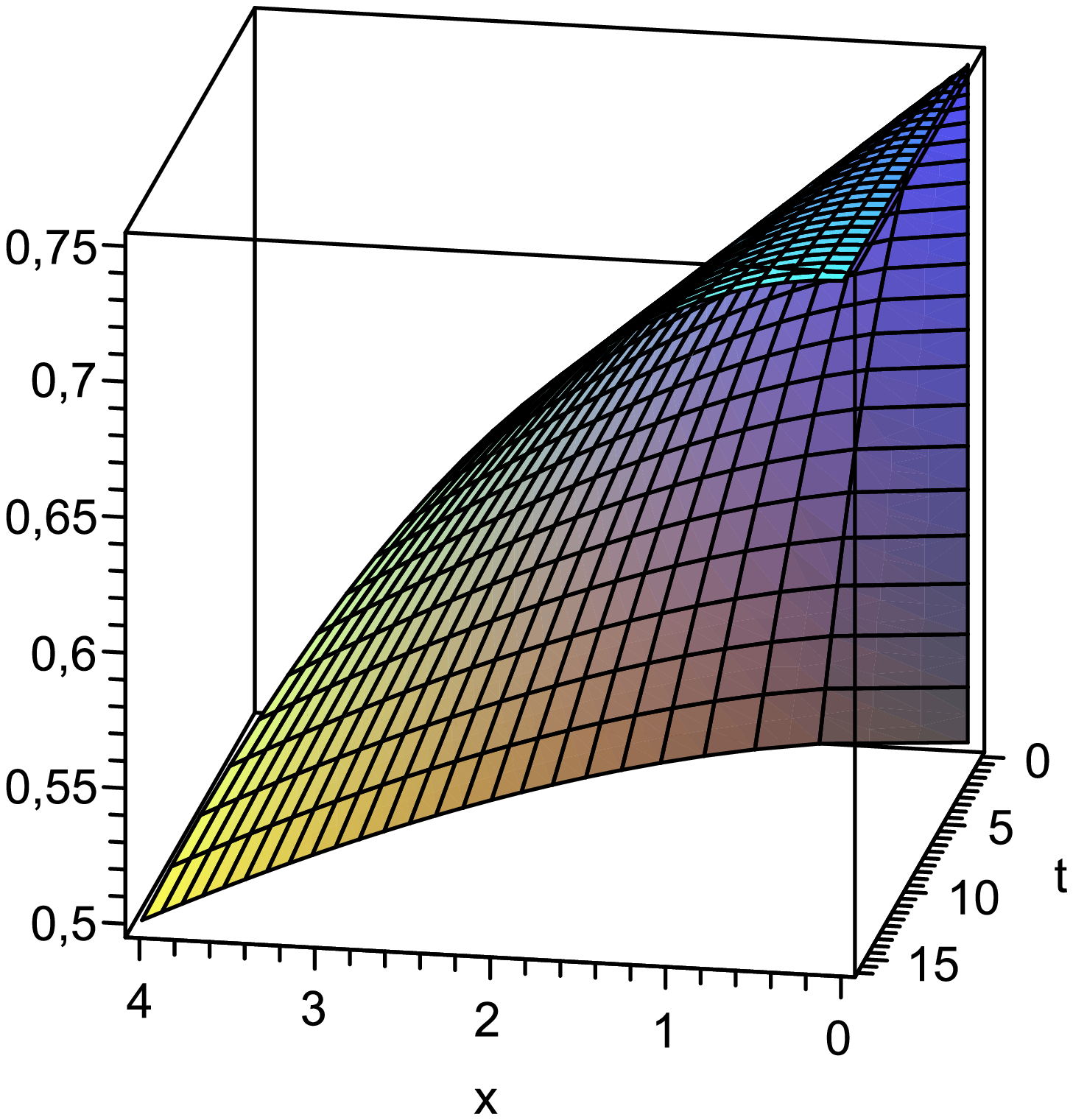}\end{center}
\end{minipage}
\hfill
\begin{minipage}{7cm}
\begin{center}\includegraphics[width=8cm]{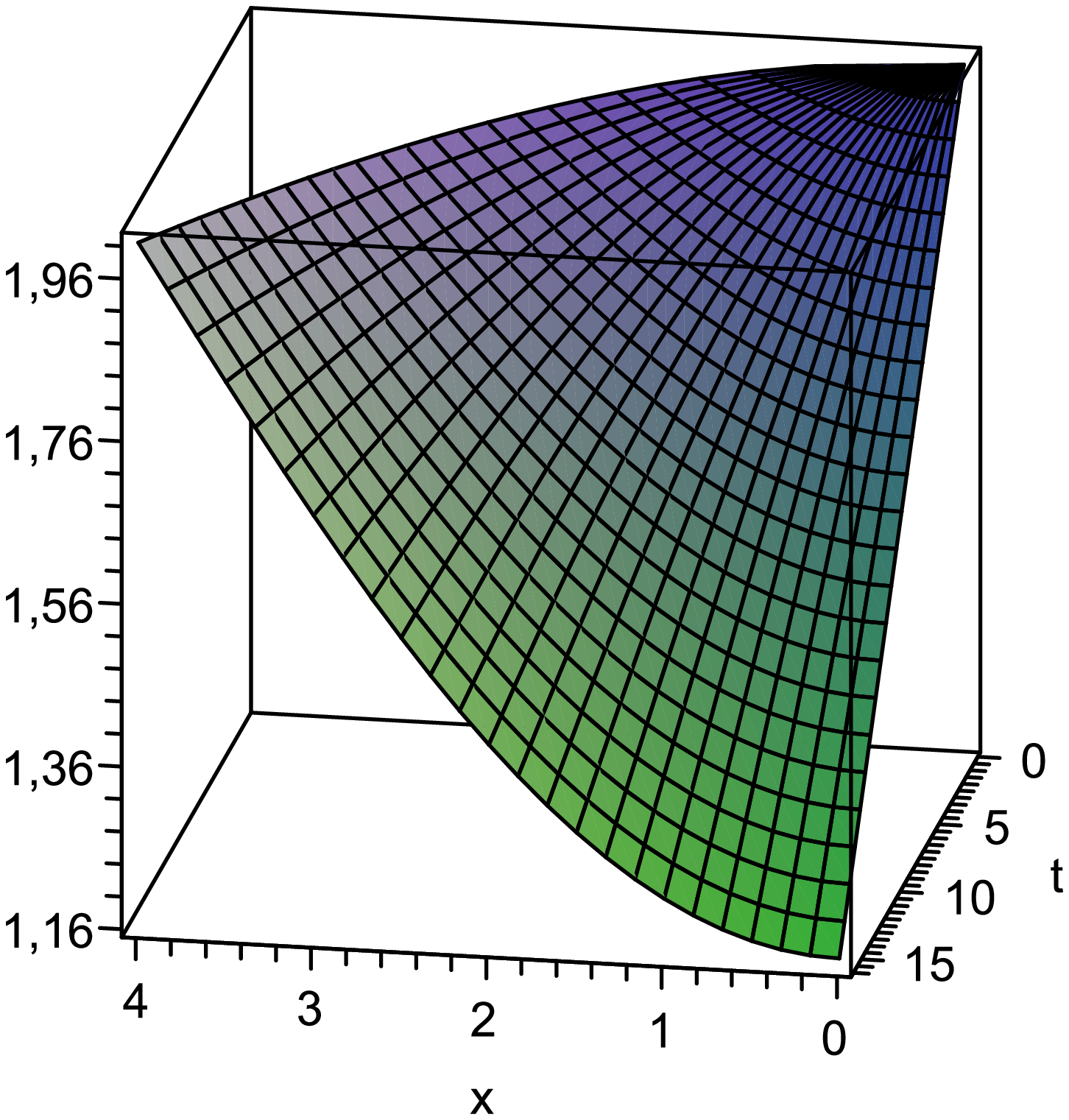}\end{center}
\end{minipage}
\caption{ Surfaces representing the concentrations $\alpha$ (left)
and $c$ (right) of the form  (\ref{5}) for parameters $m=1$,
 $\alpha_*=0.5, \ c_{\infty}=2, \ \omega_0=1, \
q_0=0.5$ } \label{f1}
\end{figure}
\begin{figure}
\begin{minipage}{7cm}
  \quad  \quad \begin{center}\includegraphics[width=8cm]{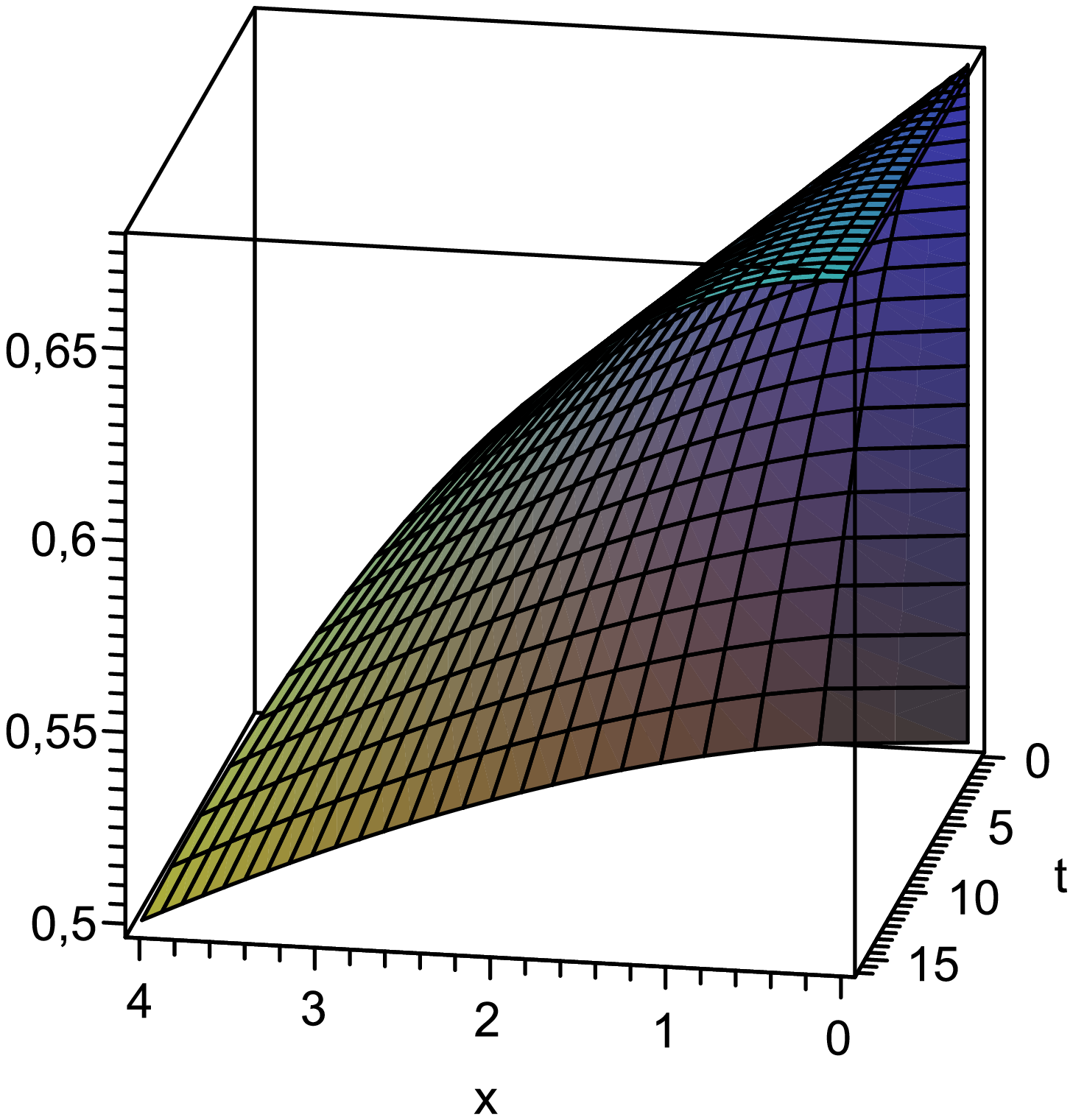}\end{center}
\end{minipage}
\hfill
\begin{minipage}{7cm}
\begin{center}\includegraphics[width=8cm]{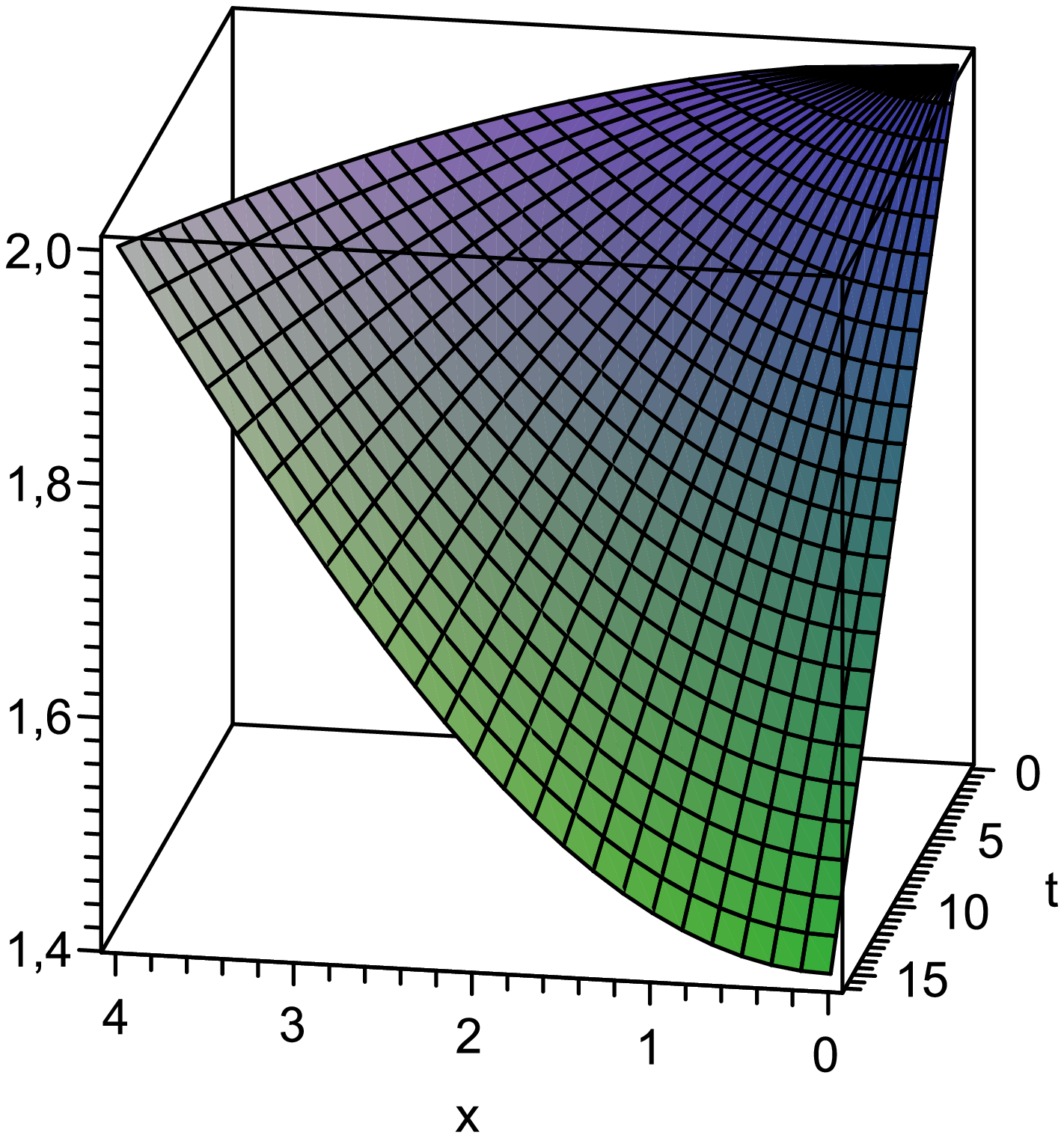}\end{center}
\end{minipage}
\caption{ Surfaces representing the concentrations $\alpha$ (left)
and $c$ (right) of the form  (\ref{5}) for parameter $m=1/2$ (other
parameters as  in Fig.\,\ref{f1}). }\label{f2}
\end{figure}
\begin{figure}
\begin{minipage}{7cm}
  \quad  \quad \begin{center}\includegraphics[width=8cm]{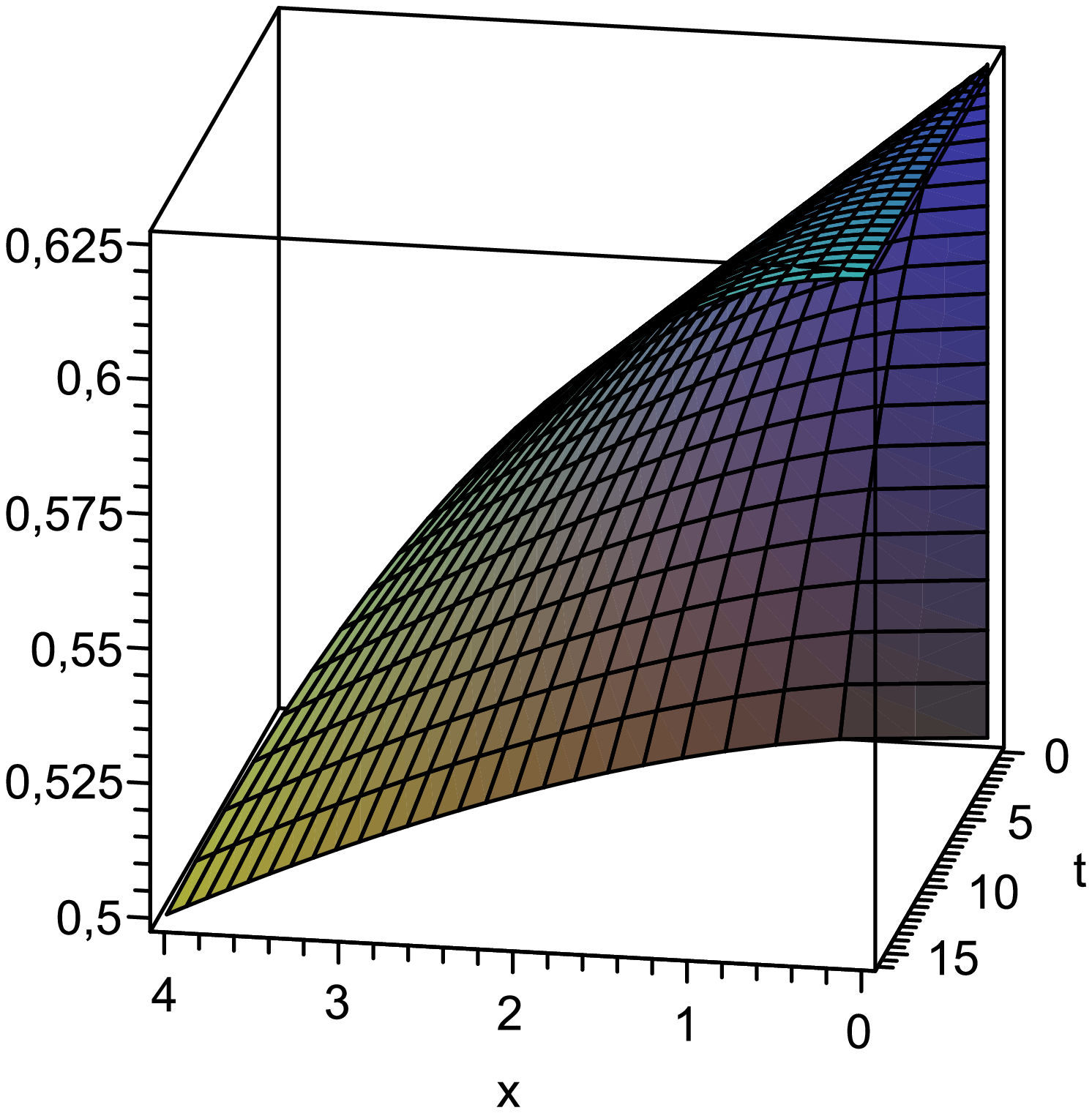}\end{center}
\end{minipage}
\hfill
\begin{minipage}{7cm}
\begin{center}\includegraphics[width=8cm]{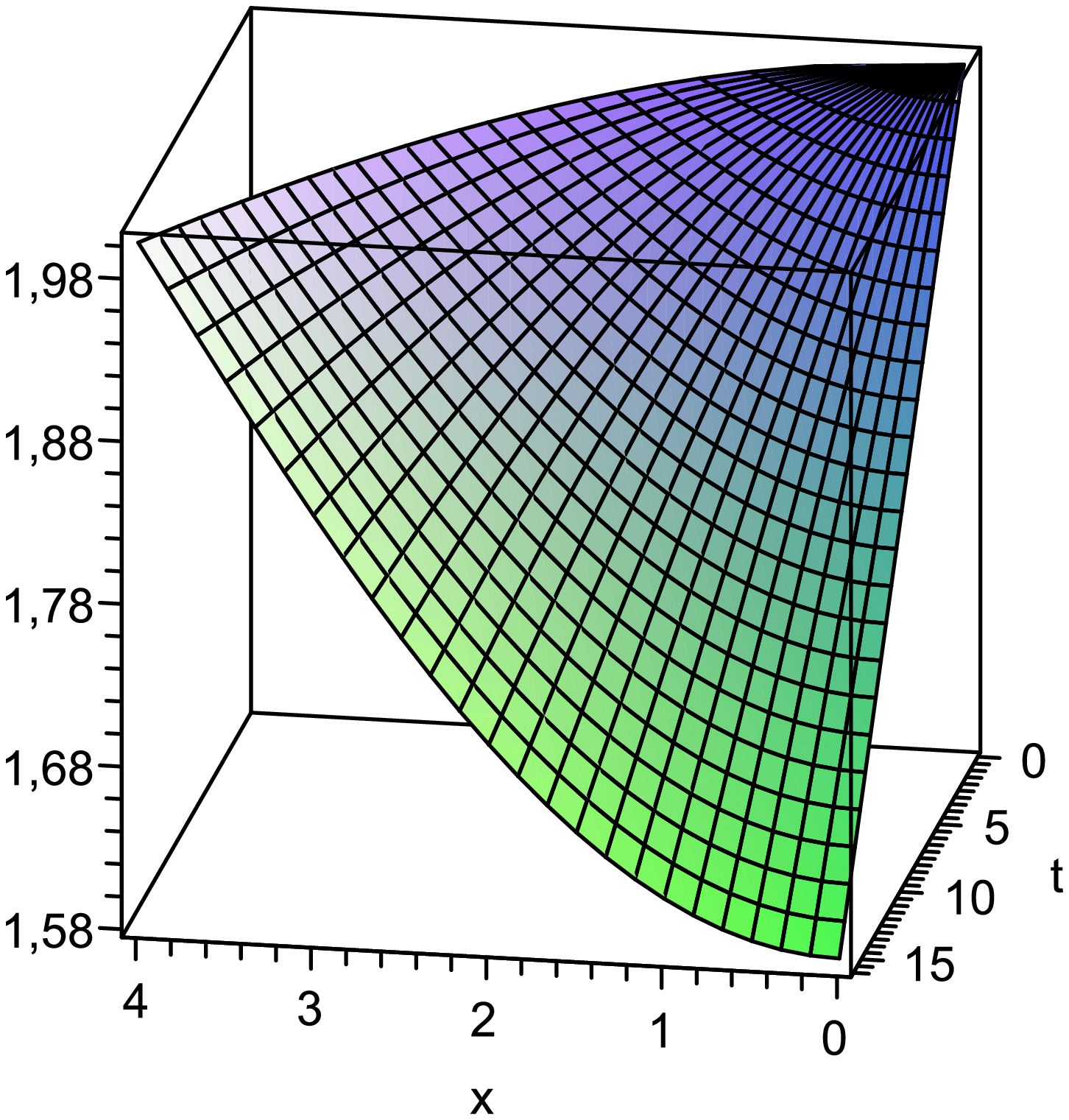}\end{center}
\end{minipage}
\caption{ Surfaces representing the concentrations $\alpha$ (left)
and $c$ (right) of the form  (\ref{5}) for parameter $m=0$ (other
parameters as  in Fig.\,\ref{f1}). }\label{f3}
\end{figure}


 \begin{remark}\label{rem-4} A similar result can be obtained for arbitrary
  diffusivity in the governing equation  (\ref{0}), i.e. the drag
  coefficient $k(\alpha)$, however, the structure of the  function $ f(\varphi)$ will essentially depend on the diffusivity.
\end{remark}

\begin{figure}
\begin{minipage}{7cm}
  \quad  \quad \begin{center}\includegraphics[width=8cm]{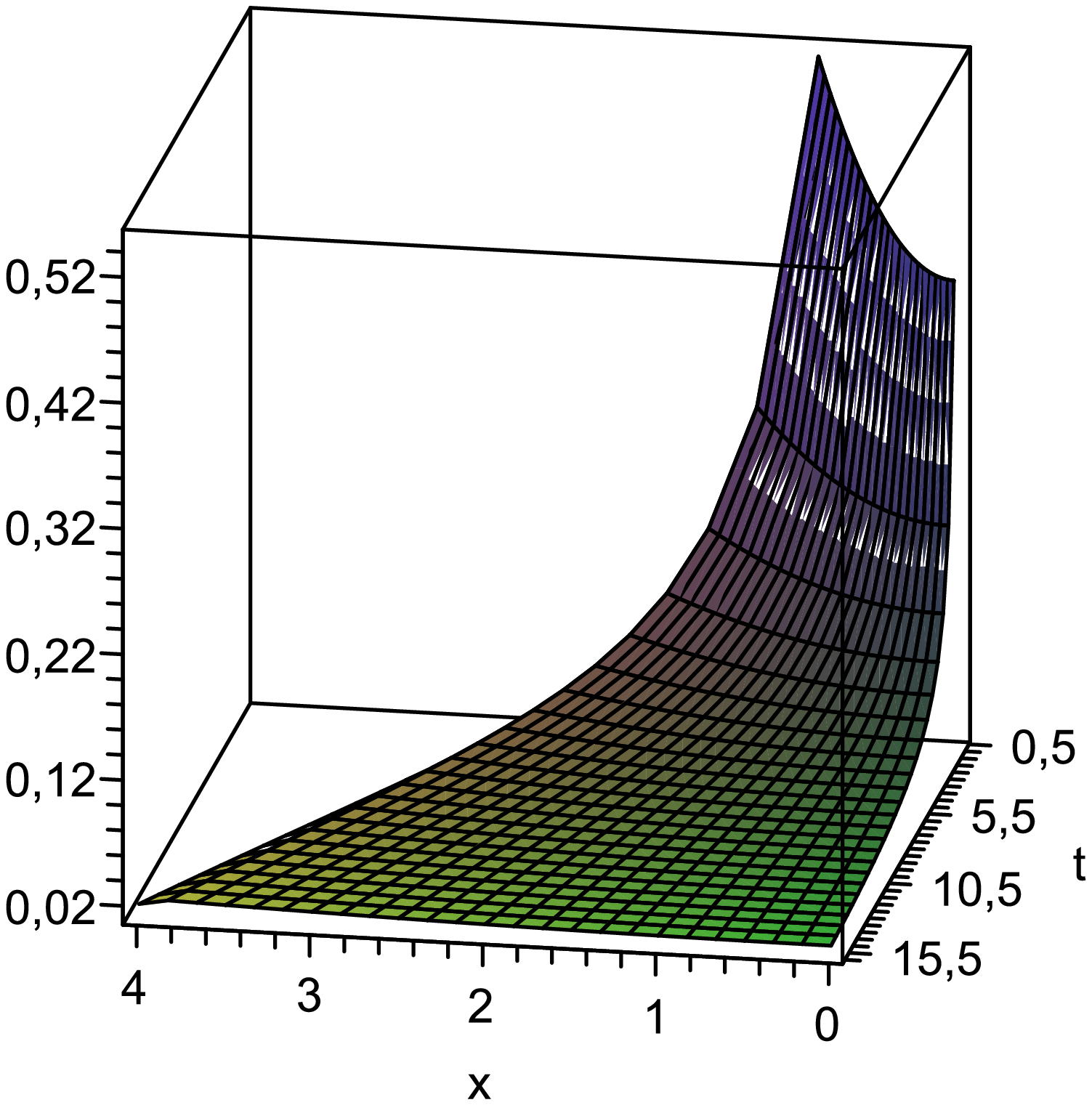}\end{center}
\end{minipage}
\hfill
\begin{minipage}{7cm}
\begin{center}\includegraphics[width=8cm]{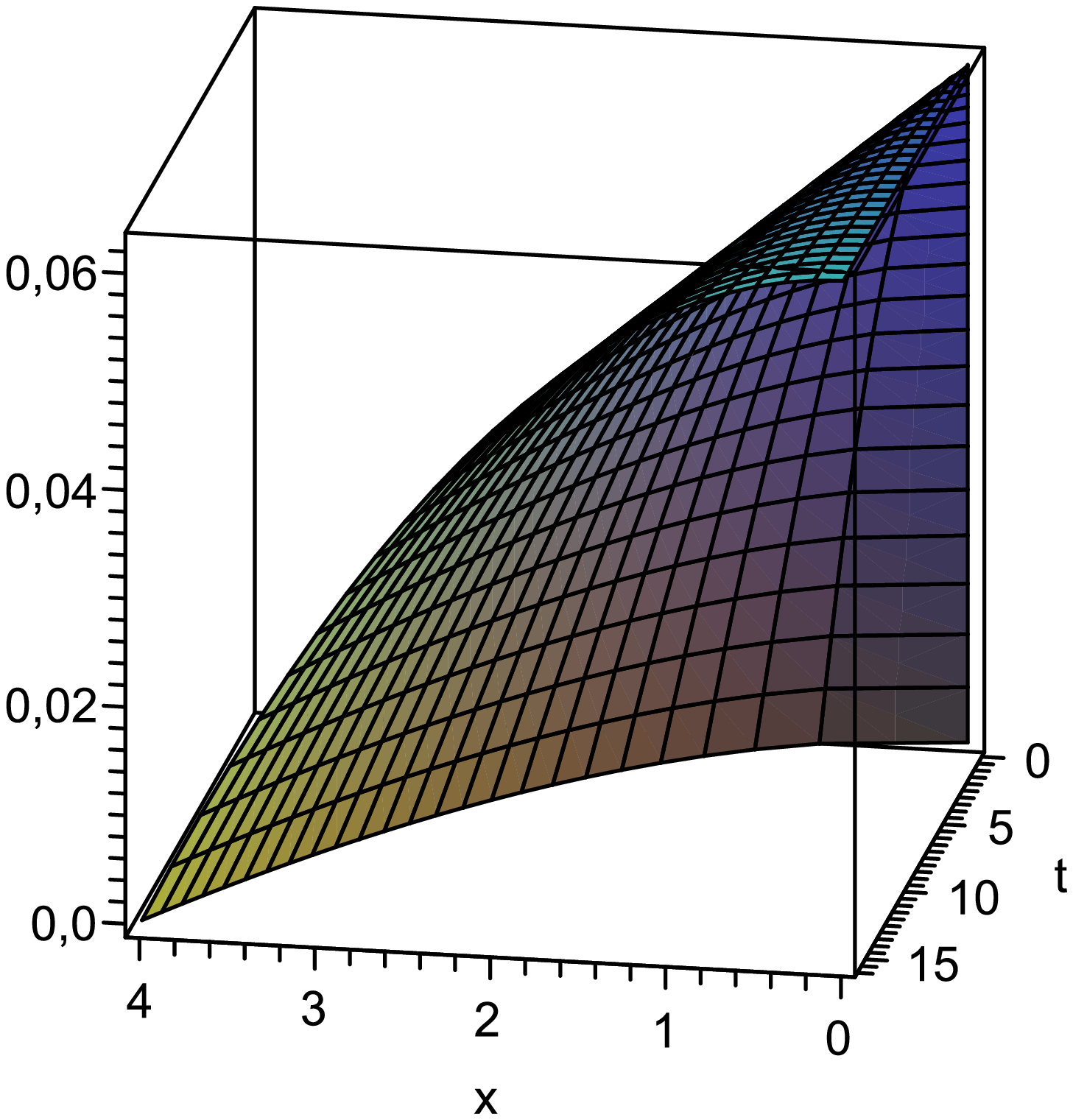}\end{center}
\end{minipage}
\caption{ Surfaces representing the function $S(\alpha(t,x),c(t,x))$
(left) and $Q(\alpha(t,x),c(t,x))$ (right) on the known
concentrations given by  (\ref{5})
(other parameters as  in Fig.\,\ref{f3}). }\label{f4}
\end{figure}

 Now we present the final formulae for
     the original concentrations of tumour cells
     $\alpha(t,x)$ and nutrient (or drug)  $c(t,x)$ using formulae  (\ref{*}), ansatz (\ref{1b})  and  solution (\ref{3}):
\begin{equation}\label{5}
 \begin{array} {l}
 \alpha(t,x)=\alpha_*+\frac{\alpha^{1-m}_*}{4}\left(\omega^2_0-\frac{x^2}{t}\right),  \ x< \omega_0\sqrt{t},\\
c(t,x)= c_{\infty} -\frac{\alpha^{1-m}_*
q_0}{48}\,t\left(\omega^2_0- \frac{x^2}{t}\right)\left(5\omega^2_0- \frac{x^2}{t}\right), \  x< \omega_0\sqrt{t},\\
R(t)= \omega_0\sqrt{t}.
 \end{array}\end{equation}

Formulae (\ref{3}) give the exact solution of the nonlinear BVP with
a moving boundary for the governing equations
\begin{equation}\label{6}
 \begin{array} {l}
 \alpha_t = \left(\alpha^m \alpha_{x}\right)_{x}+\frac{f(\alpha - \alpha_*)} {c - c_{\infty}},\\
0 = c_{xx} - q_0(\alpha - \alpha_*), \end{array}\end{equation} and
the boundary conditions
\begin{equation}\label{6a}
 \begin{array} {l}
x=0: \ \alpha_x=c_x=0,\\
x=R(t): \ \alpha = \alpha_*, \ c = c_{\infty}, \\
x=R(t): \ R'=-\frac{\alpha_x}{\alpha^{1-m}_*},
 \end{array}\end{equation}
 where the function $f$ is defined by (\ref{4}), in particular one obtains the cubic polynomial   for $m=0$:
 \begin{equation}\label{7}
 f=\frac{q_0}{3\alpha_*}(\alpha - \alpha_*)(\alpha - \alpha_1)(\alpha_2-\alpha), \ \alpha_1=\alpha_*(1-\omega^2_0), \alpha_2=\alpha_*\frac{6+\omega^2_0}{4},
 \end{equation}
hence \begin{equation}\label{8} S(\alpha, c)=
\frac{q_0}{3\alpha_*}\frac{(\alpha - \alpha_*)(\alpha -
\alpha_1)(\alpha_2-\alpha)}{c_{\infty} - c}. \end{equation}

This solution for some specified parameters is presented in
Fig.\,\ref{f1}, Fig.\,\ref{f2} and Fig.\,\ref{f3}. We note that the
positive parameter $\alpha_*<1$  can be interpreted as a natural
concentration of tumour cells,  while $c_{\infty}$ is
 the concentrations of  nutrient (drug) in the medium surrounding the tumour. The positive parameter $\omega_0$ may be found from additional biologically motivated conditions. Notably the exact solution (\ref{5}) can be easily  generalised to one with two arbitrary parameters by the time translation $t \to t+t_0 $.

It should be noted that the functions  $ f(\varphi)$  and
$g(\varphi)$  in the cases $m=0$ (constant diffusion) and $m=1$ (as
in the porous medium equation)
 are positive for the  solution  (\ref{3}), meaning that the corresponding
 functions $S(\alpha, c)$ and $Q(\alpha, c)$ (see p.570 in \cite{by-ki-2003})
 are positive.
 The cases $m\not=0, 1$  lead to a  much more complicated structure for  $ f(\varphi)$ and need
  to be examined separately for correctly-specified values of $m$.
  For example, the functions $S(\alpha, c)$ and $Q(\alpha, c)$ are
  positive on the solution for $m=1/2$ in the case of the
  parameters $\alpha_*$ and  $\omega_0$    used in   Fig.\,\ref{f2}.

\begin{figure}
\begin{minipage}{10cm}
  \quad  \quad \begin{center}\includegraphics[width=8cm]{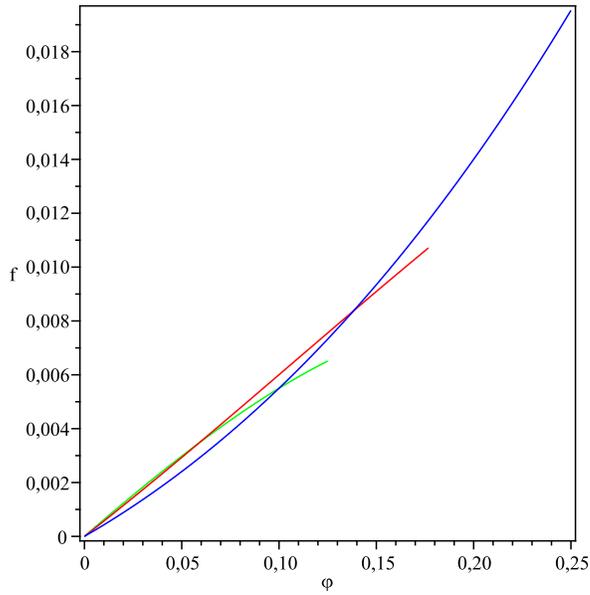}\end{center}
\end{minipage}
\caption{ Curves representing the functions $f(\varphi)$
  (\ref{4}) for parameters $m=0$ (green),
$m=0.5$ (red) and $m=1$ (blue).
The other parameters are as in  Fig.\,\ref{f1}. }\label{f5}
\end{figure}

We now turn to regimes in which the results may be susceptible to
biological application, largely  returning to the notation of
\cite{by-ki-2003}.  The effective diffusivity of the cells is given
by
\begin{equation}\label{4-2}D(\alpha)=\frac{\alpha}{k(\alpha)}(1-\alpha)^2\frac{d}{d\alpha}\left(\alpha
\Sigma\right).\end{equation} That cell-cell repulsion can be
expected to increase at high densities implies that
\begin{equation}\label{4-3}\frac{d^2}{d\alpha^2}\left(\alpha\Sigma\right)>0,\end{equation} but
this increasing contribution to (\ref{4-2}) may be offset by the
$(1-\alpha)^2$ prefactor (which itself will be less significant
because $k(1)=0$ typically holds) that in part reflects the overall
mass balance \begin{equation} \label{4-4} \alpha
v_c+(1-\alpha)v_w=0\end{equation} between cell and water velocities,
$v_c$ and $v_w$, respectively  (so when $1-\alpha$ is small so
typically is the cell velocity).  The upshot is nevertheless that
cases with $m\geq0$ are those of most relevance, with increasing $m$
corresponding to stronger cell-cell repulsion at high densities.

$S(\alpha,c)$ will be positive when $c$ represents a nutrient,
with $S$ an increasing function of $c$ in both cases (corresponding
to
 $\beta>0$ in the above). In this case, $S$ will be an increasing function of $\alpha$
for sufficiently small $\alpha-\alpha_*$ ( and linear in $\alpha$)
  and $S=0$ at $\alpha=\alpha_*$.
One may note that the curves  in Fig.~\ref{f5} satisfy these
properties. In fact, all the curves are increasing functions,  the
green and red curves in Fig.~\ref{f5} behave  as  linear functions,
while the blue curve is linear for sufficiently small
$\varphi=\alpha - \alpha_* $ (the curves are built on the interval
$[0, \varphi_{max}]$ with $\varphi_{max}$ prescribed by the first
formula in (\ref{3})). Moreover,  $f(\varphi)=0$ for $\varphi=0$
because the free  parameter $\omega_0$ was taken to be $1$ (see
$\alpha_1$ in (\ref{7})). Notably, $S$ may decrease again for larger
$\alpha$, reflecting the limited supply of water to make new cells.
The function $S$ defined by (\ref{8}) possesses such property
provided $\alpha > \alpha_{max} \in (\alpha_*,\alpha_2)$ , where
$\alpha_{max} $ is a local maximum of $f$ from (\ref{7}) and  can be
easily calculated.


In the drug case, $S$ would typically be taken positive for small
$c$ and negative for large $c$ (with $S$ a decreasing function of
$c$) --- this behaviour cannot be captured by the above
nonlinearities, so in interpretation of the results, $c$ should be
regarded as everywhere sufficiently large that $S$ is negative,
again with $\beta>0$. Various dependencies on $\alpha$ can then be
identified, and even   $S=0$ at $\alpha=0$ need not hold  (though
additional constraints need imposing if $\alpha$ reaches  zero).

$Q$ should be positive in both interpretations (absorption, rather
than generation, though the generation of toxic materials in the
drug case could allow $Q<0$ to be
contemplated) and an increasing function of $c$ and $\alpha$.
According to Theorem~\ref{th-3}  $Q=g(U)V^{-\beta+1}$ (with $U$ and
$V$ given in (\ref{*})), therefore one needs $\beta>1$ and an
increasing function $g$ in order to satisfy the above  requirement.


$Q>0$ implies by the maximum principle that $c$ will be an
increasing function of $x$, having a minimum at $x=0$. Such
behaviour of $c$ is pictured in Fig.~\ref{f1}--\ref{f3}  for
different values of $m$. The qualitative form of $\alpha$ will
depend on the initial data and the form of $S$; in the case $S>0$,
the nutrient supply is highest at the edge of the tumour, but
$\alpha$ is held at $\alpha_*$ there by the stress balance (more
water is taken up to prevent $\alpha$ exceeding $\alpha_*$ due to
cell division), so, perhaps counterintuitively,  it is quite
possible for $\alpha$ to be maximal at $x=0$ if $S$ remains
sufficiently large, despite the decrease in $c$. However, such
behaviour is more natural for $S<0$ (the drug case) since the
cell-kill rate can be expected to be highest at the edge of the
tumour.

\section{
Higher-dimensional tumour growth model without cell viscosity}

In Section 4, we examined the one-dimensional model (\ref{4-1a})
 describing the tumour growth.  Here we generalise the model to higher
  dimensions under assumption that tumour cells create a radially
  symmetrical shape (a circular cylinder for $n=1$ and a sphere for $n=2$). In this case, the nonlinear BVP
   (\ref{4-1a}) takes the following form:
 \begin{equation}\label{5-1a}
 \begin{array} {l}
 U_t = \frac{1}{r^n}\left(r^n(U+\alpha_*)^mU_{r}\right)_{r}+S(U+\alpha_*,c_{\infty}-V),\\
0 = \frac{1}{r^n}\left(r^nV_{r}\right)_{r} + Q(U+\alpha_*,c_{\infty}-V),\\
r=0: \ U_r=V_r=0,\\
r=R(t): \ U=V=0,\\
r=R(t): \ R'=- \alpha^{m-1}_* U_r,
 \end{array}\end{equation}
 where $n=1,2$ and the same notation are used as in Section 4, however,
 the space variable $x$ is replaced by the variable $r=\sqrt{x_1^2+\dots+x_n^2+x_{n+1}^2}$.

 One may check that the governing equations of this  BVP with $S=f(U)V^{-\beta}$ and   $Q=g(U)V^{-\beta+1}$ are  invariant w.r.t.
the  Lie symmetry operator
 \begin{equation}\label{5-1c} X=2\beta t\partial_t+\beta r\partial_r + 2V\partial_V,\end{equation}
 which has the same structure as (\ref{1c}).

  \begin{theo}\label{th-4}
  The nonlinear  BVP (\ref{5-1a}) with $S=f(U)V^{-\beta}$ and   $Q=g(U)V^{-\beta+1}$  is invariant with respect to the two-dimensional MAI generated by the Lie symmetry operators  (\ref{5-1c}) and $\partial_t$.
\end{theo}

  Thus, the Lie symmetry  (\ref{5-1c})  generates the ansatz
 \begin{equation}\label{5-1b}
 \begin{array} {l}
 U = \varphi(\omega), \ \omega=\frac{r}{\sqrt{t}}\,,\\
V = t^{\frac{1}{\beta}}\psi(\omega),
 \end{array}\end{equation}
which immediately specifies the function
 $R(t)=\omega_0\sqrt{t}$\, and  reduces the two-dimensional  BVP (\ref{5-1a})
 (with $S=f(U)V^{-\beta}$ and   $Q= g(U)V^{-\beta+1}$)  to the one-dimensional problem
 \begin{equation}\label{5-2}
 \begin{array} {l}
(\varphi+\alpha_*)^m\varphi''+m(\varphi+\alpha_*)^{m-1}\left(\varphi'\right)^2+
\frac{n}{\omega}\,(\varphi+\alpha_*)^m\varphi'+\frac{\omega}{2}\,\varphi'+f(\varphi)\psi^{-\beta}=0,\\
\psi''+\frac{n}{\omega}\,\psi'+g(\varphi)\psi^{-\beta+1}=0,\\
\omega=0: \ \varphi'=\psi'=0,\\
\omega=\omega_0: \ \varphi=\psi=0, \
\varphi'=-\frac{\alpha^{1-m}_*\omega_0}{2}.
 \end{array}\end{equation}

 The nonlinear problem (\ref{5-2}) is still not integrable for arbitrary given coefficients, however,  one with   $\beta=-1$ is again a solvable case. Direct calculations show that formulae (\ref{3}) and (\ref{4}) can be generalised in this case as follows
 \begin{equation}\label{5-3}
 \begin{array} {l}
 \varphi=\frac{\alpha^{1-m}_*}{4}\left(\omega^2_0-\omega^2\right),  \ \omega< \omega_0,\\
\psi= \frac{\alpha^{1-m}_*
q_0}{16(n+3)}\left(\omega^2_0-\omega^2\right)\left(\frac{n+5}{n+1}\omega^2_0-\omega^2\right),
\  \omega< \omega_0
 \end{array}\end{equation}
and
 \begin{equation}\label{5-4}
 \begin{array} {l}
 g= q_0\varphi,  \quad  q_0>0, \\
f= \frac{q_0\alpha^{m-1}_*}{n+3}\varphi\left(\varphi+\frac{\alpha^{1-m}_*\omega^2_0}{n+1}\right)\\
\quad  \times \Bigg(
(m+\frac{n+1}{2})\alpha^{1-m}_*(\varphi+\alpha_*)^m-
\frac{m\alpha^{2-m}_*}{4}\left(\alpha^{-m}_*\omega^2_0+4\right)(\varphi+\alpha_*)^{m-1}
-\varphi+\frac{\alpha^{1-m}_*}{4}\omega^2_0\Bigg).
 \end{array}
\end{equation}

In order to obtain the final formulae for the original
concentrations of tumour cells $\alpha(t,x)$ and nutrient (drug)
$c(t,x)$, we  use formulae  (\ref{*}), ansatz (\ref{5-1b}) and
solution (\ref{5-3}). The formulae obtained coincide with those
presented in  (\ref{5}) for $\alpha(t,x)$ and $R(t)$ while for the
nutrient (drug) concentration we have the solution
\begin{equation}\label{5-5} c(t,r)= c_{\infty} -\frac{\alpha^{1-m}_*
q_0}{16(n+3)}\,t\left(\omega^2_0-
\frac{r^2}{t}\right)\left(\frac{n+5}{n+1}\omega^2_0-
\frac{r^2}{t}\right), \ r< \omega_0\sqrt{t}.
 \end{equation}

The biomedical interpretation of the various regimes  coincides with
that in Section 4 and is omitted  here.

\section{ Conclusions }

The main part of this paper  is devoted to the Lie symmetry classification
of the class of parabolic-elliptic  systems (\ref{1-3}). First of all,
 the structure of
 form-preserving (admissible) transformations for the  systems of the form
(\ref{1-3}) is established (see Theorem~\ref{th-1}) in order to
establish possible  relations between systems that  admit equivalent
MAIs. Theorem~\ref{th-2} gives a complete list  of inequivalent
parabolic-elliptic  systems
consisting of 35 systems, which are invariant under non-trivial
MAIs, i.e. under the three- and higher-dimensional Lie algebras.
Moreover, we have established all possible subclasses of
parabolic-elliptic  systems (see Tables~\ref{tab-3} and~\ref{tab-4})
with non-trivial Lie symmetry, which are reducible to those
 listed in Tables~\ref{tab-1} and~\ref{tab-2} by the relevant form-preserving
 transformations. Notably,  our result is a new confirmation of importance
 of finding form-preserving transformations because they allow to reduce essentially
  (in contrary to the group of  equivalence transformations) the number of inequivalent
  systems. In fact, the traditional Lie--Ovsiannikov algorithm, which is  based on  the group
  of  equivalence transformations, would lead to 57 parabolic-elliptic  systems instead of 35
   systems listed in Tables~\ref{tab-1} and~\ref{tab-2} because 22 systems from Tables~\ref{tab-3} and~\ref{tab-4} should be added.

From the applicability point of view, the most interesting parabolic-elliptic
 systems occur in Table~\ref{tab-1} because the systems from Table~\ref{tab-2} are semi-coupled. It is
  worth to note that MAIs for the systems arising in all the cases excepting 7 and
   14 are some analogs of those for single reaction-diffusion equations. In fact,
   fixing the diffusion coefficient in the first equation  of the corresponding system
    one may easily identify its generic analog in Table 4 \cite{ch-se-ra-08} (see cases 3--5 and 7 therein).
    For example, MAIs of the systems from cases 8 and 9 contain
 the well-known algebra $sl(2,\mathbb{R} )$ as a subalgebra (see
 the second, third and fourth operators in the last column). The
 same subalgebra occurs for  the diffusion equation with the same
 diffusivity (see case 3 in Table 4 \cite{ch-se-ra-08}). By the way, Ovsiannikov was
 the first who identified this special case for nonlinear diffusion equations \cite{ovs-1959}.
  Of course, the systems in Table~\ref{tab-1} have more general structures (many of them contain arbitrary
   functions as coefficients) in contrast to their analogs listed in Table 4 \cite{ch-se-ra-08}.
    However, it  happens on the regular basis if one considers systems instead of single equations
     (see, e.g., case 10 in Table 1 \cite{ch-king4} involving the same diffusivity).

Cases 7 and 14 in Table~\ref{tab-1}, have no analogs among single
reaction-diffusion equations. Moreover, the relevant MAIs  generate
infinite-dimensional Lie algebras because they contain
 the operators $X^{\infty}$ involving an arbitrary smooth function $\varphi(t)$ (see the last
  column of Table~\ref{tab-1}). Interestingly that the both MAIs from Cases 7 and 14 contain again
 the well-known algebra $sl(2,\mathbb{R} )$ as a subalgebra. In fact,
  if one sets  $\varphi(t)=t$ and $\varphi(t)=t^2$ then the operators $\partial_t$, $X^{\infty}$ with $\varphi(t)=t$ and $X^{\infty}$ with
$\varphi(t)=t^2$ produce nothing else but $sl(2,\mathbb{R} )$ with
new representations (it is a simple task to check the commutator
relations), which do not happen in the case of single
reaction-diffusion equations.

In order to demonstrate applicability of our pure theoretical result,   we apply the  Lie symmetry corresponding to one-parameter group of scale transformation
for solving  a  (1+1)-dimensional  BVP  with a free boundary modeling tumour growth \cite{by-ki-2003}.
We reduce the given nonlinear  BVP to that governed by ODEs. As a result,
 its exact solution  was constructed under  additional restrictions on the coefficients arising in the governing equations. Moreover, possible  biological
interpretations of the solution  is discussed. In particular, it was shown that the results obtained allow plausible  interpretation in  the case when the model describes   growth  of tumour cells  consuming a nutrient.
Finally, the results for  the above (1+1)-dimensional  BVP  were generalised on
multi-dimensional case  under assumption that tumour cells create a radially
  symmetrical shape.

\section {Acknowledgments}
This research was supported by a Marie Curie International Incoming
Fellowship to the first author within the 7th European Community
Framework Programme (project BVPsymmetry 912563).


\begin{thebibliography}{99}

\footnotesize


\bibitem {ames} Ames, W.F.: Nonlinear partial differential
equations in engineering. Academic Press, New York (1972)

\bibitem {arr-15} Arrigo, D.J. Symmetries Analysis of Differential Equations --- An Introduction.
Wiley, USA (2015)

\bibitem{bl-anco} Bluman G.W., Anco C.: Symmetry and Integration Methods for Differential Equations. Springer, New York (2002)

 \bibitem{b-k}   Bluman, G.W.,  Kumei, S.:   Symmetries and differential
 equations. Springer,  Berlin (1989)

 \bibitem{by-ki-2003}   Byrne, H., King, J.R.,  McElwain, D.L.S.,
 Preziosi, L.:
 A two-phase model of solid tumour growth. Appl. Math. Letters.
 \textbf{16},
 567--573 (2003)


 \bibitem{che-dav2014}  Cherniha, R.,  Davydovych, V.: Reaction-diffusion systems with constant diffusivities:
  conditional symmetries and form-preserving transformations. In Algebra, Geometry and Mathematical
   Physics, Springer Berlin Heidelberg \textbf{85},  533--553 (2014)

\bibitem {ch-king}  Cherniha, R., King, J.R.:
 Lie    symmetries of nonlinear  multidimensional
reaction-dfiffusion systems:I.  J. Phys. A: Math. Gen.  \textbf{33},
267--282 (2000)

\bibitem {ch-king1}  Cherniha, R., King, J.R.:
 Addendum: Lie    symmetries of nonlinear  multidimensional
reaction-dfiffusion systems:I.  J. Phys. A: Math. Gen. \textbf{33},
7839--7841 (2000)

\bibitem {ch-king2} Cherniha, R., King, J.R.:
  Lie    symmetries of nonlinear  multidimensional
reaction-dfiffusion systems: II.  J. Phys. A: Math. Gen.
\textbf{36}, 405--425 (2003)

 \bibitem {ch-king4}  Cherniha, R., King, J.R.: Nonlinear reaction-diffusion systems  with variable diffusivities:
 Lie symmetries, ans\"atze and exact solutions.  J. Math. Anal. Appl.
 \textbf{308}, 11--35 ( 2005)

 \bibitem {ch-king06} Cherniha, R., King, J.R.:  Lie symmetries
  and conservation laws of non-linear multidimensional reaction--diffusion systems
  with variable diffusivities. IMA J. Appl. Math. \textbf{71}, 391--408
  (2006)

 \bibitem{ch-kov11a} Cherniha, R.,  Kovalenko, S.:   Lie symmetries and reductions
  of multi-dimensional boundary value problems
  of the Stefan type. J. Phys. A: Math. Theor. \textbf{44}, 485202 (25 pp.) (2011)


\bibitem {ch-se-98} Cherniha, R.,  Serov, M.:
Symmetries, ans\"atze  and exact solutions of  nonlinear
second-order evolution equations with convection term.
  Euro. J. Appl. Math. \textbf{9}, 527--542 (1998)

\bibitem{ch-se-ra-08} Cherniha, R.,  Serov, M.,  Rassokha, I.: Lie symmetries and
form–preserving transformations of reaction—diffusion—convection
equations. J. Math. Anal. Appl.  \textbf{342} , 1363—-1379 (2008)


  \bibitem{fss}  Fushchych, W.I.,   Shtelen, W.M.,   Serov, M.I.:
   Symmetry analysis and exact solutions of equations of nonlinear
mathematical physics, Kluwer (1993)

\bibitem{winternitz1992}  Gazeau,  J.P.,  Winternitz, P.: Symmetries  of
 variable  coefficient  Korteweg-de  Vries  equations. J.  Math.  Phys.
 \textbf{33},  4087--4102 (1992)

    \bibitem {kingston-91}   Kingston, J.G.:
     On point transformations of evolution equations.  J. Phys. A: Math. Gen.
     \textbf{24},
     L769--L774 (1991)

  \bibitem {kingston-98} Kingston, J.G.,  Sophocleous, C.:
On form-preserving point transformations of partial differential
equations.  J. Phys. A: Math. Gen.   \textbf{31}, 1597--1619 (1998)


\bibitem {ibrag-94} Knyazeva, I.V., Popov, M.D.:
A system of two diffusion equations. In CRC Handbook of Lie Group
Analysis of Differential Equations, Boca Raton, CRC Press
\textbf{1}, 171--176 (1994)

\bibitem {lotka}  Lotka, A.J.:
Undamped oscillations derived from the law of mass action. J. Amer.
Chem. Soc. \textbf{42}, 1595 -- 1599 (1920)


\bibitem {mur2} Murray, J.D.: Mathematical biology.  Springer, Berlin (1989)

\bibitem {mur2003} Murray, J.D.: Mathematical biology  II: Spatial
 Models and Biomedical Applications. Springer, Berlin (2003)

\bibitem {niederer-78} Niederer, U.: 1978. Schr\"odinger invariant generalised heat equation. Helv.
Phys. Acta. \textbf{51}, 220--239 (1978)


 \bibitem {okubo} Okubo, A., Levin, S.A.: Diffusion and ecological problems.
Modern perspectives, 2-nd ed. Springer, Berlin (2001)

\bibitem {olv}   Olver, P.:
 Applications of Lie groups to differential equations, Springer,
 Berlin (1986)

\bibitem{ovs-1959}  Ovsiannikov, L.V.:  Group relations of the equation of non-linear heat conductivity.
            Dokl. Akad. Nauk SSSR   \textbf{125},  492--495 (1959)


 \bibitem {tor-tra-15}  Torrisi, M.,  Tracina, R.:
   An application of equivalence transformations to reaction diffusion equations. Symmetry
   \textbf{7},
   1929--1944 (2015)

    \bibitem {turing}  Turing, A.M.: The chemical basis of
    morphogenesis,  Phil. Trans. R. Soc. B \textbf{237}, 37--72 (1952)


\bibitem {volterra} Volterra, V.: Variazionie fluttuazioni del numero
d`individui in specie animali conviventi. Mem. Acad. Lincei.
\textbf{2},  31--113 (1926)

 \bibitem {zulehner-ames} Zulehner, W.,   Ames, W.F.: Group analysis of
a semilinear vector diffusion equation. Nonlinear Anal. \textbf{7},
945--969 (1983)



















\end{thebibliography}
\end{document}